\documentclass[prb,aps,onecolumn,showpacs,amsmath,amssymb]{revtex4-2}% 
\usepackage{graphicx}% Include figure files
\usepackage{dcolumn}% Align table columns on decimal point
\usepackage{lipsum}
\usepackage[dvipsnames]{xcolor}
\usepackage{ulem}
\usepackage[percent]{overpic}
\usepackage{gensymb}
\usepackage{color}
\usepackage{comment}
\usepackage{amsmath}
\usepackage{amsfonts}
\usepackage{amssymb}

\usepackage{epstopdf}

\begin{document}

%\begin{frontmatter}

%\title{Tuning properties of strongly correlated materials by means of strong laser pulses}% Force line breaks with \\
% \title{Post-Floquet engineering of electronic phase transitions by intense laser pulses}
%\title{Attosecond multidimensional spectroscopy and control of strongly correlated systems}

\title{Sub-cycle multidimensional spectroscopy of  strongly correlated materials}

\author{V. Valmispild}
\affiliation{Institute of Theoretical Physics, University of Hamburg, Notkestraße   9, 22607 Hamburg, Germany}
%\affiliation{The Hamburg Centre for Ultrafast Imaging, Luruper Chaussee 149, 22761 Hamburg, Germany}
\affiliation{European XFEL, Holzkoppel 4, 22869 Schenefeld, Germany}

\author{E. Gorelov} 
\affiliation{European XFEL, Holzkoppel 4, 22869 Schenefeld, Germany}

\author{M. Eckstein} 
\affiliation{Department of Physics, University of Erlangen-Nuremberg, 91058 Erlangen, Germany}

\author{A.~I. Lichtenstein}
\affiliation{Institute of Theoretical Physics, University of Hamburg, Notkestraße 9, 22607 Hamburg, Germany}
\affiliation{The Hamburg Centre for Ultrafast Imaging, Luruper Chaussee 149, 22761 Hamburg, Germany}
\affiliation{European XFEL, Holzkoppel 4, 22869 Schenefeld, Germany}

\author{H. Aoki} 
\affiliation{Department of Physics, The University of Tokyo, Hongo, Tokyo 113-0033, Japan}
\affiliation{National Institute of Advanced Industrial Science and Technology (AIST), Tsukuba 305-8568, Japan}

\author{M.~I. Katsnelson} 
\affiliation{Institute for Molecules and Materials, Radboud University, Heyendaalseweg 135, 6525AJ Nijmegen, The Netherlands}

\author{M. Ivanov} 
\affiliation{Max-Born-Institut, Max-Born-Str. 2A, 12489 Berlin, Germany,\\ Institute of Physics, Humboldt University of Berlin, Newtonstraße 15, 12489 Berlin, \\ Department of Physics, Imperial College London, SW7 2BW London, United Kingdom}

\author{O. Smirnova} 
\affiliation{Max-Born-Institut, Max-Born-Str. 2A, 12489 Berlin, Germany,\\ Technical University Berlin, Straße des 17. Juni 135, 10623 Berlin, Germany,\\ Department of Physics, The Solid State Institute Building, Israel Institute of Technology, Haifa, Israel}

\date{\today}

\begin{abstract}
Strongly correlated solids are extremely complex and fascinating quantum systems, where new states continue to emerge, especially when interaction with light triggers interplay between them. In this interplay, sub-laser-cycle electron response is particularly attractive as a tool for ultrafast manipulation of matter at PHz scale. 
Here we introduce a new type of non-linear multidimensional spectroscopy, which allows us 
to unravel the sub-cycle dynamics of strongly correlated systems interacting with  few-cycle infrared pulses and the complex interplay between different correlated states evolving on the sub-femtosecond time-scale. We demonstrate that single particle sub-cycle electronic response is extremely sensitive to correlated many-body dynamics and provides direct access to many body response functions.
For the two-dimensional Hubbard model under the influence of ultra-short, intense electric field transients, we demonstrate that our approach can resolve pathways of charge and energy flow between localized and delocalized many-body states on the sub-cycle time scale and follow the creation of 
a highly correlated state surviving  after the end of the laser pulse. 
Our findings open a way towards a regime of imaging and manipulating strongly correlated materials at optical rates, beyond the multi-cycle approach employed in Floquet engineering,  with the sub-cycle response being a key tool for accessing many body phenomena.
\end{abstract}

\maketitle

%\section{Main Text}
The advent of attosecond pulses \cite{kienberger2004atomic,sansone2006isolated,ferrari2010high}, attosecond spectroscopy \cite{itatani2002attosecond,cavalieri2007attosecond,calegari2014ultrafast,nisoli2017attosecond,seiffert2017attosecond,pazourek2015attosecond,dahlstrom2012introduction,vampa2015linking,silva2018high,silva2019topological}, and lightwave electronics\cite{goulielmakis2007attosecond,jimenez2020lightwave,jimenez2021sub}, which aim to resolve and control light-driven electron motion on sub-laser cycle time scales \cite{krausz2009attosecond,baltuvska2003attosecond,wittmann2009single,goulielmakis2008single},  has challenged our perception of reactivity -- a capacity of %a substance, such as an 
atoms, molecules or solids to undergo changes triggered by an external agent. 
In chemistry, it prompted a quest for charge directed reactivity, a chemical change driven by  attosecond electron dynamics \cite{Cederbaum1999,Remacle1999,Lunnemann2008,Breidbach2003,Remacle2006,Kuleff2014,Calegari2016}. Similar concept should exist in solids,
%condensed matter systems, 
where strong electron-electron correlations can lead to a rich variety of phase transitions and the appearance of new states of matter\cite{kuroda2017evidence,yin2018giant,bradlyn_science_2016} due to electronic response to intense light\cite{topp2018all}.

So far, investigation of light-driven changes in strongly correlated materials was focused on the Floquet regime\cite{Oka_Aoki_2009,oka_review_Floquet,Mikami_2016}, giving rise to the powerful concept of Floquet engineering of quantum materials, which typically focuses on laser-cycle-averaged modifications of material properties.
Yet, in strongly correlated systems the sub-laser-cycle time scale is also highly relevant: a typical electron-electron interaction parameter $U\sim 1 eV$ corresponds to the time-scale $\Delta t\sim 1/U \sim$ 1 fs, well below the cycle of a standard infrared driver, with the respective 
dynamics potentially leading to such remarkable features as e.g. a transition from Coulomb repulsion to effective electron-electron attraction induced by half-cycle pulses \cite{tsuji2012repulsion,tsuji2011dynamical}.

One way to probe and control the sub-laser-cycle electronic response is to use few-cycle pulses with controlled carrier-envelope phase (CEP) \cite{krausz2009attosecond,wittmann2009single,goulielmakis2008single,schultze2010delay}. In solids, these pulses have been used to detect  photoemission delays \cite{ossiander2018absolute} and quantify the time-scale of non-linear response to light \cite{sommer2016attosecond}, image  surface states in topological insulators \cite{schmid2021tunable}, resolve and control highly non-linear electronic response in bulk dielectrics, 2D materials, and nano-structures \cite{ludwig2020sub,putnam2017optical,higuchi2017light,seiffert2017attosecond,lakhotia2020laser}. Yet, the physical picture of electron-electron correlations evolving on the \textit{sub-cycle} scale in strongly correlated systems remains elusive.

Here we introduce a 
sub-cycle multidimensional spectroscopy of electron dynamics in solids and apply it to a strongly correlated system. 
Our approach uses the CEP-dependence of the correlated multi-electron response 
%CEP-dependent response to reveal and decode  correlated electron dynamics accompanying metal to insulator transition, driven by a few-cycle CEP-controlled pulse. 
to 
%demonstrate the key role of the sub-cycle electronic response in strongly correlated solids and 
decode the complex interplay between different many-body states, triggered by the 
interaction with a few-cycle mid-IR control pulse. 
%ME
%The analysis of the multi-dimensional spectra  has the potential to uncover the physical picture of the correlated electron dynamics in space and time.
Ultimately, the  analysis of the multi-dimensional spectra allows us to uncover  the physical picture of the underlying correlated dynamics in this system, both in space and time. 
Fundamentally, we show that the sub-cycle one-particle response is able to track  many-body dynamics by providing direct access to many-body response functions. The possibility to spectroscopically analyze the underlying excitation pathways, as introduced below, is  key to understanding non-thermal materials control.

We consider a half-filled Hubbard model 
on the two-dimensional square 
lattice for fermions, supporting  a realistic two-dimensional band dispersion with the characteristic  
van Hove singularity and sharp band edges.  The lattice is driven by a strong field linearly-polarized along the lattice diagonal, triggering  a fully two-dimensional response (in contrast to previously employed Bethe- or hypercube-lattices \cite{joura2015long} or one-dimensional chains \cite{silva2018high,aron2012dimensional}.)
%will result in a situation similar to the Bethe-lattice or infinite-dimensional hypercube.  
To treat the non-perturbative 
time-dependent problem, we employ the non-equilibrium 
extension \cite{schmidt2002nonequilibrium,aoki2014nonequilibrium} of 
the dynamical mean-field theory (DMFT) \cite{georges1996dynamical}.
The algorithm and its realization are described in Ref.~\cite{eckstein2010interaction} (see Methods). The method was benchmarked against exact diagonalization of one-dimensional finite 
chain\cite{silva2018high}, see  Supplementary Information (SI). 
The implementation is based on the NESSi simulation package for non-equilibrium Green's functions\cite{schuler2020nessi}.
We employ a single-band Hubbard model 
%and adopt the hopping 
with parameters for undoped 
La$_2$CuO$_4$ (LCO):  the lattice constant $a_{0}=3.78 \AA$, the nearest-neighbour hopping 
$T_1=0.43$~eV \cite{markiewicz2005one}, the Hubbard $U=2.5$ eV\cite{delannoy2009low}.
We use few-cycle pulses  
with central 
wavelength of $\lambda=1500$~nm (with $\omega=0.827$~eV) and
duration of 7.7 fs (full width at half-maximum, FWHM), with a 
total simulation time 32.8 fs. 
To demonstrate that our results are typical for the low-frequency regime $\omega\ll U$, 
we also present simulations for $\lambda=3000$~nm ($\omega = 0.413$ eV).  

\begin{figure}[h!]
\begin{minipage}[h]{0.49\linewidth}
\begin{overpic}[width=1\textwidth]{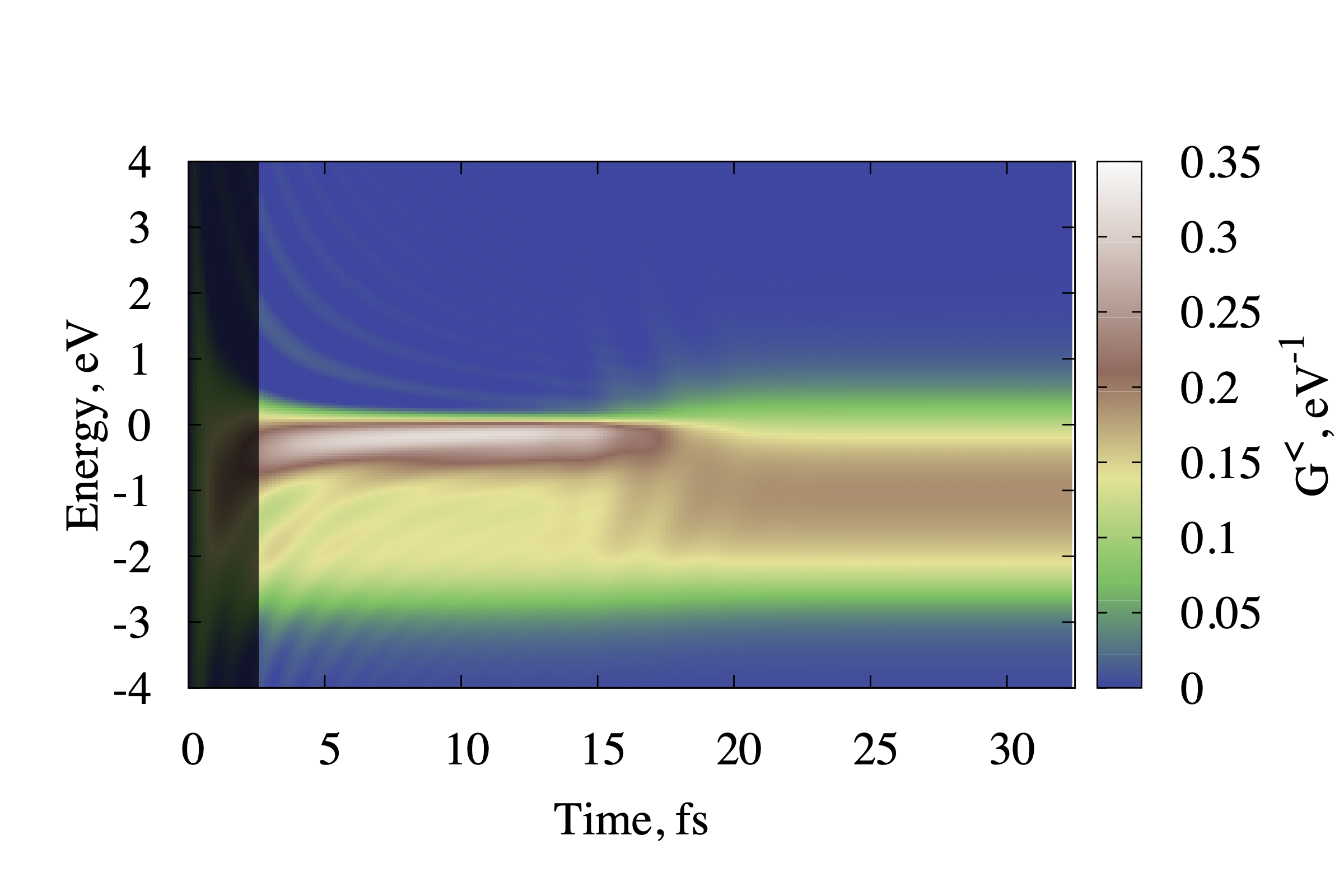}
 \put (14.6,50.5) {\textcolor{white}{(a)}}
\end{overpic}
\end{minipage}
\hfill
\begin{minipage}[h]{0.49\linewidth}
%\begin{overpic}[width=1\textwidth]{figures_/Fig_1/1b.eps}
\begin{overpic}[width=1\textwidth]{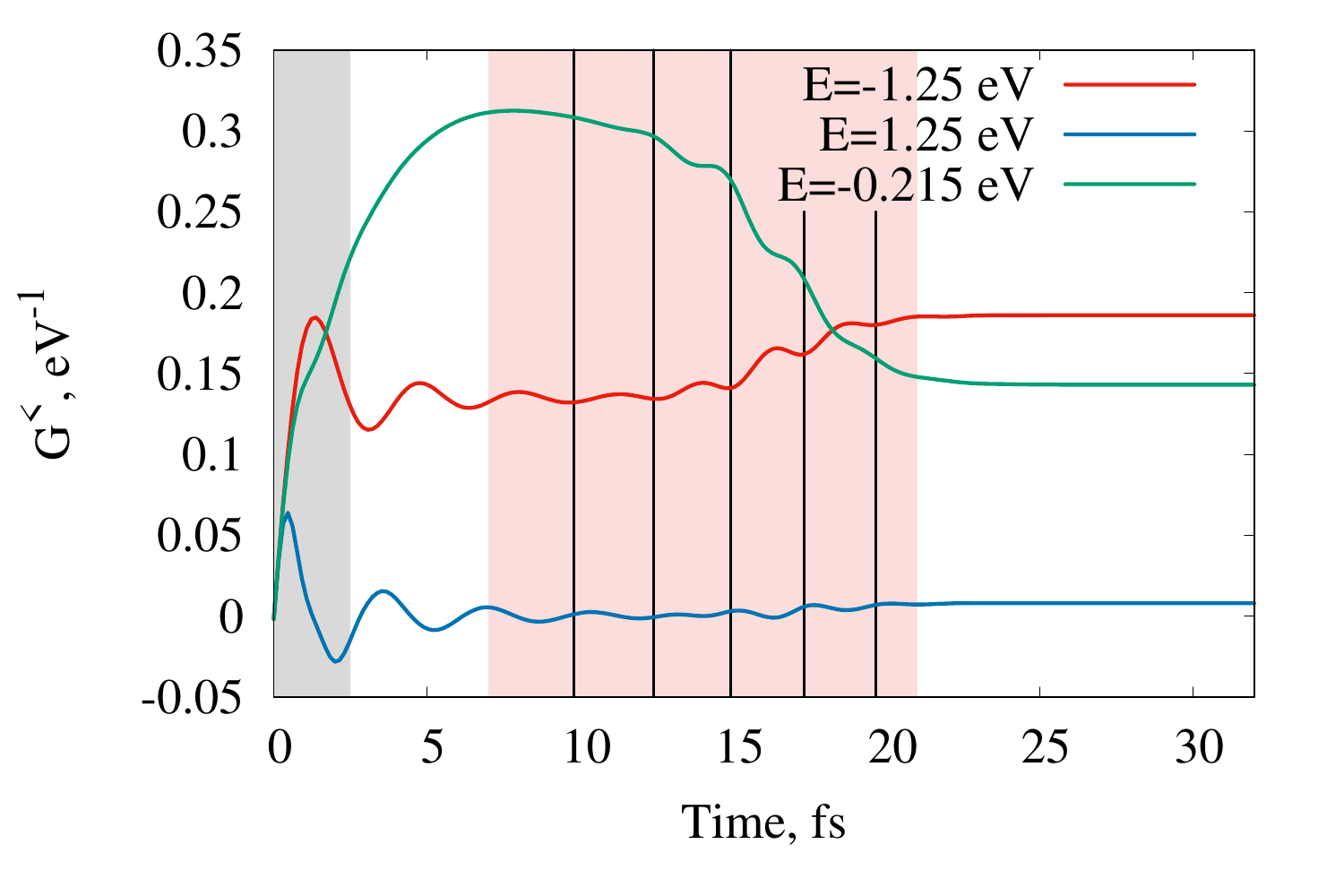}

 \put (21,58) {\textcolor{black}{(b)}}
\end{overpic}
\end{minipage}
\begin{minipage}[h]{0.49\linewidth}
\begin{overpic}[width=1\textwidth]{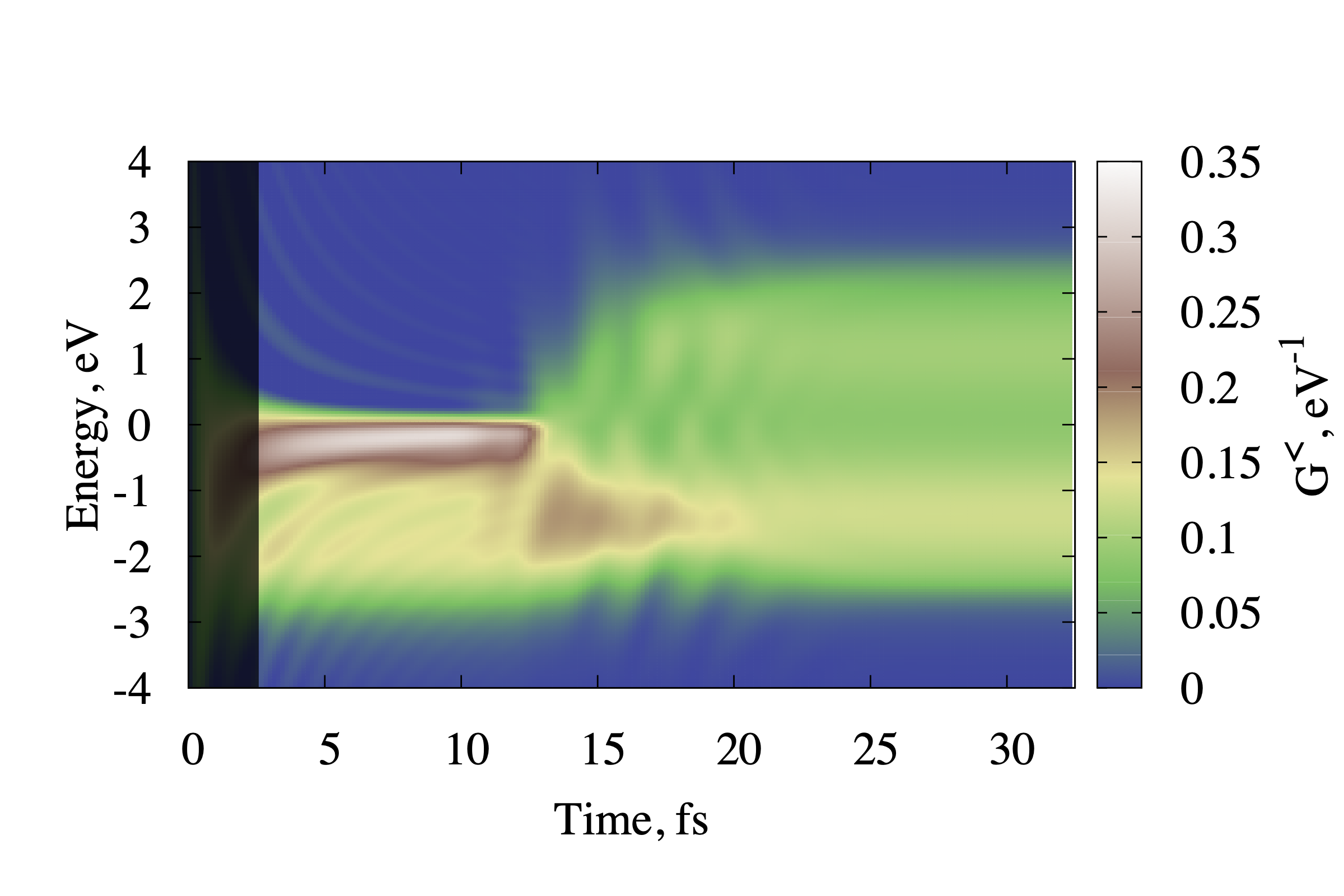}
 \put (14.6,50.5) {\textcolor{white}{(c)}}
\end{overpic}
\end{minipage}
\hfill
\begin{minipage}[h]{0.49\linewidth}
\begin{overpic}[width=1\textwidth]{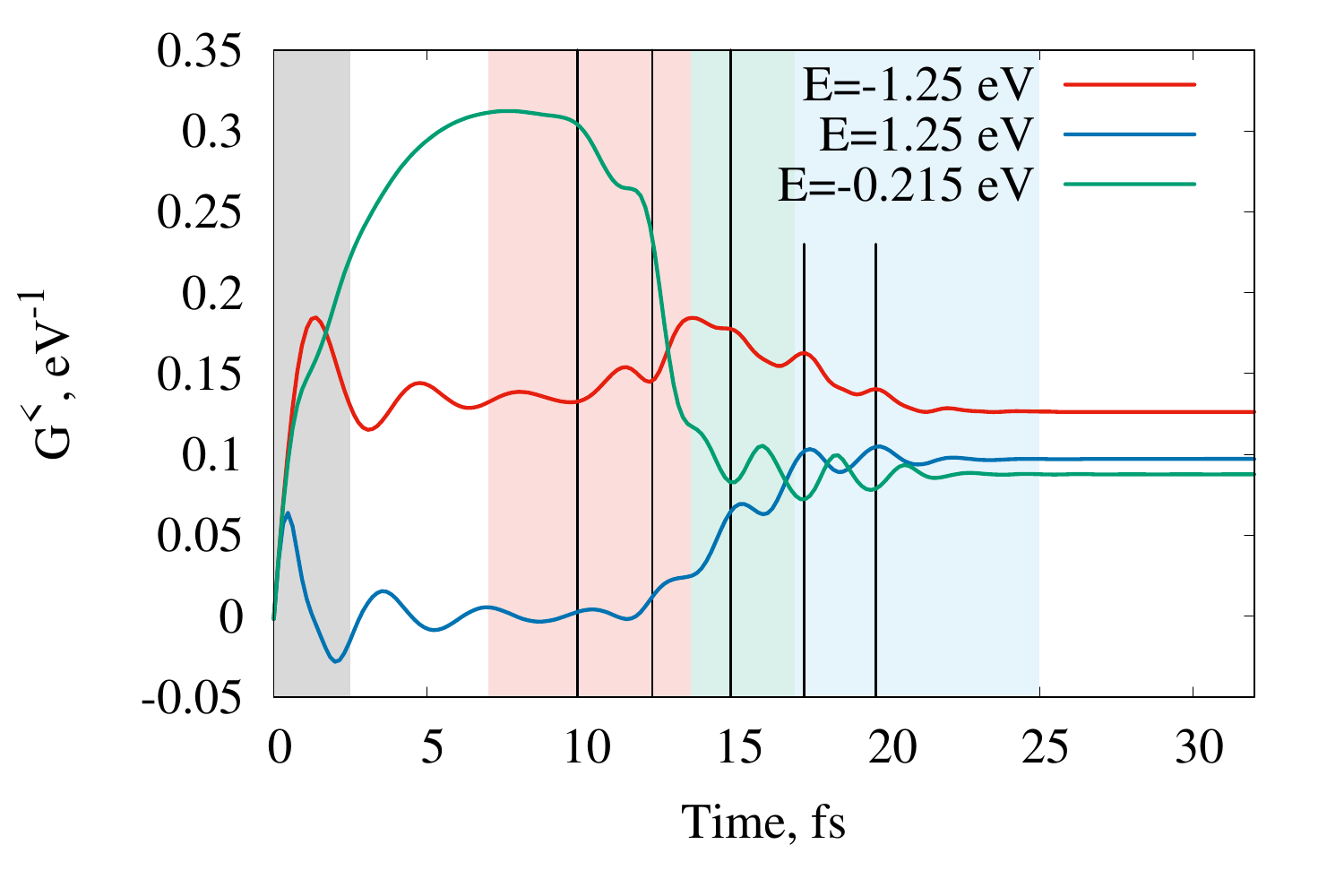}
 \put (21,58) {\textcolor{black}{(d)}}
\end{overpic}
\end{minipage}
\begin{minipage}[h]{0.49\linewidth}
\begin{overpic}[width=1\textwidth]{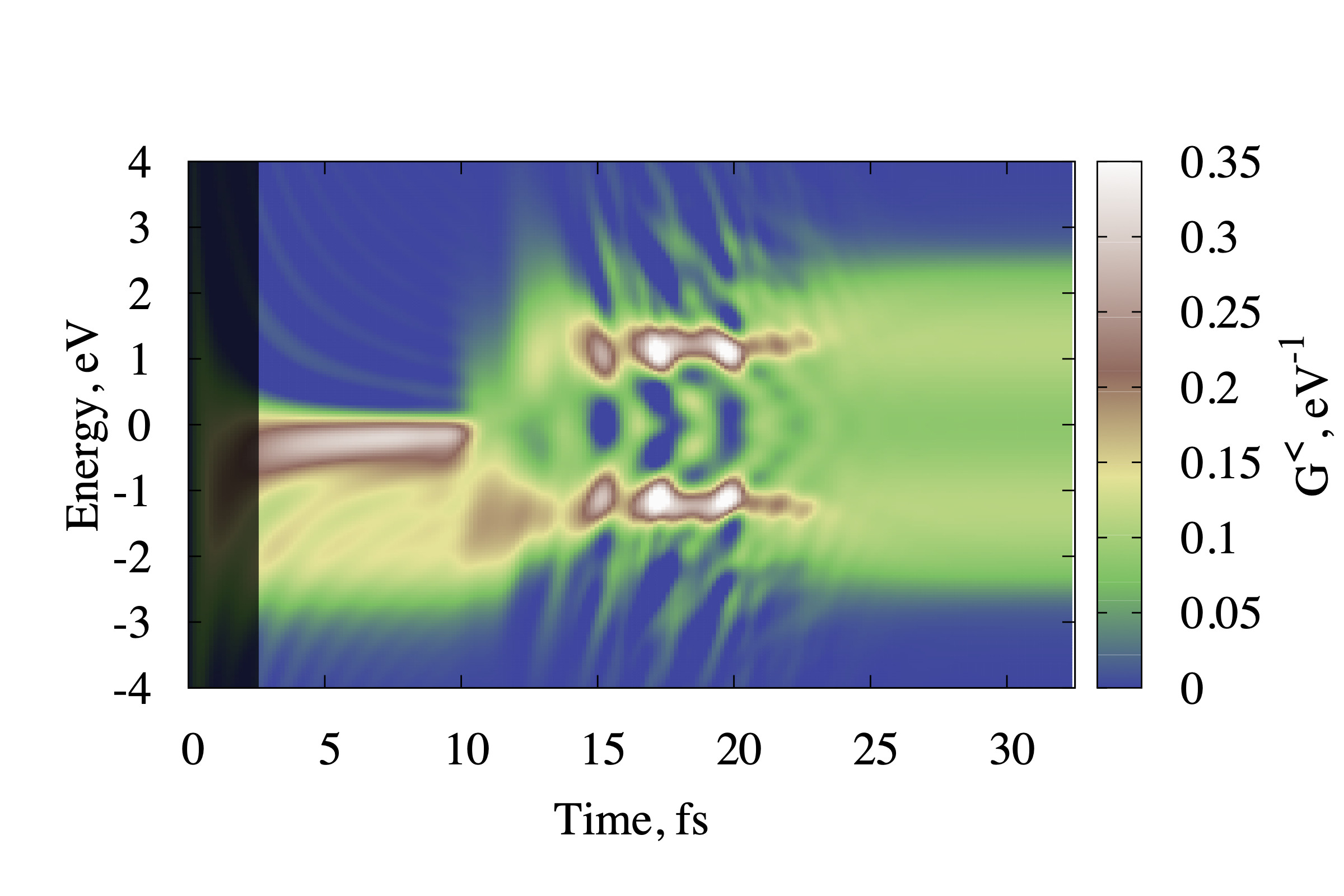}
 \put (14.6,50.5) {\textcolor{white}{(e)}}
\end{overpic}
\end{minipage}
\hfill
\begin{minipage}[h]{0.49\linewidth}
\begin{overpic}[width=1\textwidth]{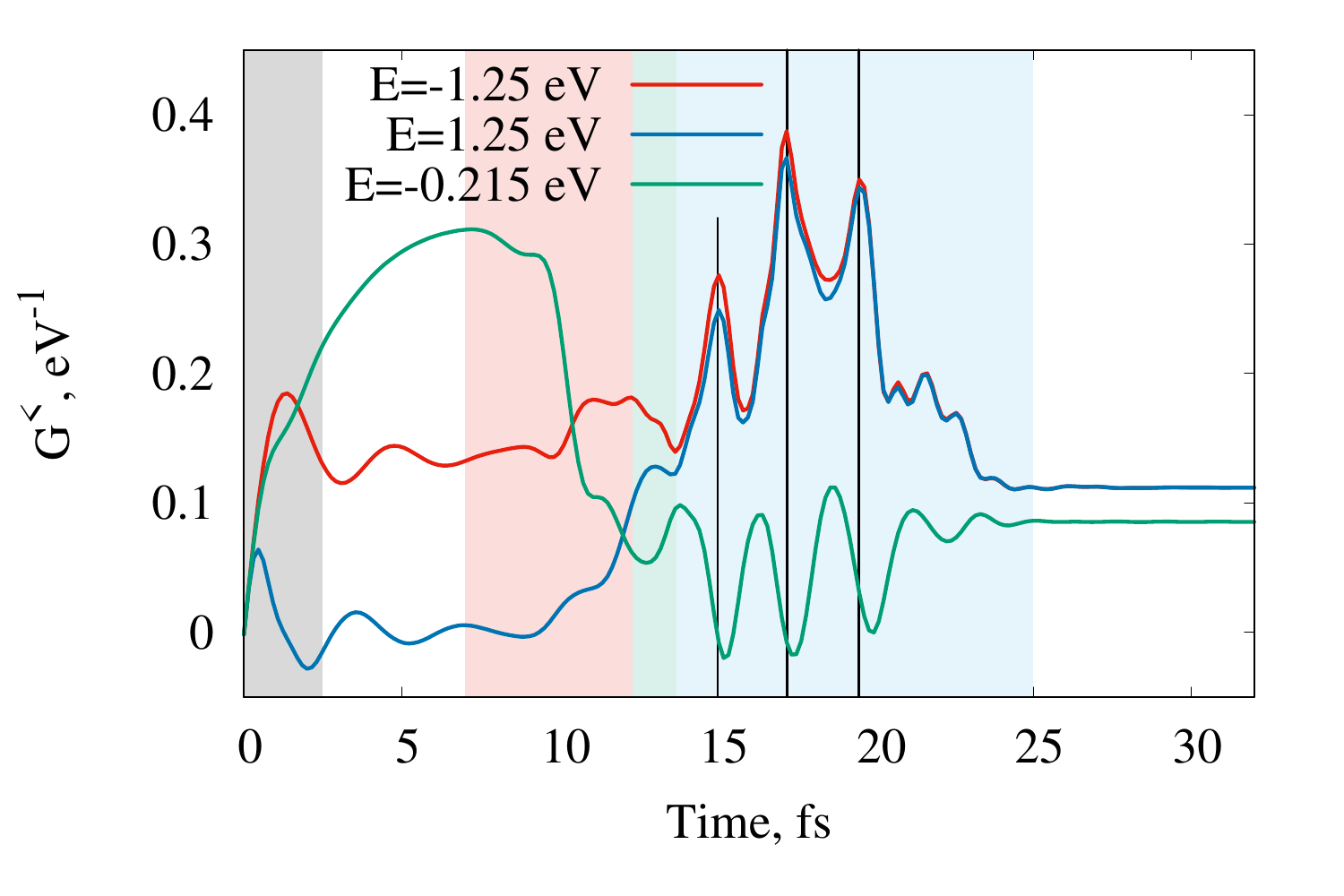}
 \put (19,58) {\textcolor{black}{(f)}}
\end{overpic}
\end{minipage}
\caption{Temporal evolution of density of states showing light-induced transition from metallic to Mott-insulating states. (a)  $F_0$=0.1 V/A, (c) 0.5 V/A,(e) 2.0 V/A. Artifact of the Fourier transform which appears at 0-2.5 fs is covered by shadow. (b,d,f) Oscillations  of electron density at energies corresponding to $LHB$  ($E=-1.25$ eV, red), $UHB$ ($E=1.25$ eV, blue), and $QP$ ($E=-0.215$ eV, green) for $F_0=0.1$ V/A (b),  $F_0=0.5$ V/A (d) and $F_0=2$ V/A (f). Red, green and blue shading marks three different regimes of field-driven dynamics. Red shading: density at $LHB$ and $UHB$ oscillates out of phase; blue shading: locking regime, density at $LHB$ and $UHB$ oscillates in phase; green shading: intermediate regime. Vertical lines indicate the maxima in $LHB$ and $UHB$ populations, correlated to minima in $QP$ populations in locking regime. }
\label{Fig1}
\end{figure}

Figures \ref{Fig1}(a,c,e) show the temporal profile of the occupied density of states (see %Eq.~\ref{Gles} in
Methods) for 
the field strengths $F_0$ from 0.1 to 2.0 V/A.
The voltage across a 
unit cell approaches the hopping rate, $a_0F_0 \sim T_1$, at  
$F_0\sim 0.1$ V/A ($I_0\sim 1.6 \times 10^{11}$W/cm$^2$). Thus, 
$F_0\sim 0.1$ V/A could  
modify the effective hopping rate within
the laser cycle and alter the structure of the correlated system. 

Indeed, the transfer of spectral weight from the quasi-particle peak (QP), 
%(located at energy $-0.215 eV$) 
located near zero energy to 
the Hubbard bands becomes prominent 
as soon as $F_0$ approaches 0.1 V/A, see Fig.~\ref{Fig1}(a): 
after the transition at $\sim 17.5$ fs, 
the spectral density remains predominantly 
in the lower Hubbard band and does not return back to the QP after 
the pulse.
Already for this field,
%lowest field strength, 
Fig.~\ref{Fig1}(a) hints at the importance of the sub-cycle response: 
the 
%ME standard 
cycle-averaged renormalized hopping $T_1\rightarrow T_1 J_0(F_0a_0/\omega)=0.97T_1$ does not lead to 
any substantial changes in the spectral density, let alone to the major 
restructuring observed  in Fig.~\ref{Fig1}(a)
(see SI).

At higher fields (Fig.~\ref{Fig1}(c,e)), 
we see substantial  transfer of the spectral density
to the 
upper Hubbard band (situated at $E=1.25$ eV), with the electron density 
peaked at the energies corresponding to the upper  and lower Hubbard bands. Crucially, this dichotomic  structure
survives well after the end of the pulse. 
Figures \ref{Fig1}(a,c,e) thus signify 
transition from a metallic to 
%ME
a highly correlated state in which the light-driven current is fully quenched (see Fig.\ref{Fig4}(d) below.)
%bad metallic electron liquid induced by the laser excitation.
%an insulating 
%%%(or bad metallic) 
%behaviour, 
%induced by the laser excitation %and stabilized by correlation.
%\smallskip

To understand these complex many-body dynamics,  we first look at 
the cuts (Figure \ref{Fig1}(b,d,f)) of the electron density for specific energies corresponding to  the lower Hubbard band (LHB, $E=-1.25$ eV), upper Hubbard band (UHB, $E=1.25$ eV) and $QP$ (maximizing at $-0.215$ eV). The exchange of population in Fig. \ref{Fig1}(d) has 
three distinct regimes, marked as three shaded areas: around 4-11 fsec (red), 
11-16 fsec (green), and beyond 16 fsec (blue). The first regime (red shading) 
shows decreasing electron density at the energy corresponding to the 
$QP$ peak and increasing density at the energies corresponding to  $LHB$  and $UHB$ 
(see also Figure \ref{Fig1}(b)), with the  populations at $LHB$ and $UHB$ energies oscillating 
out of phase (see also Figure \ref{Fig1}(b)). In the second 
regime (green shaded area in Figure \ref{Fig1}(d)), the density at $UHB$ energy raises, while the density at $LHB$ energy decreases. The third regime is most surprising: we observe in-phase oscillations of the electron density at $UHB$ and $LHB$ energies  (blue shaded area in Figure \ref{Fig1}(d)). At higher intensities the electron densities at $UHB$ and $LHB$ energies  become locked: both populations are equal and oscillate exactly in phase (blue shaded area 
at $t>14$ fs in Figure \ref{Fig1}(f)).  The maxima of these locked populations are synchronized with the minima in the density located at the $QP$. 
%  and are correlated with the instantaneous zeros of the laser field (vertical solid lines). The  minima in electron densities at $LHB$ (red) and $UHB$ (blue) energies occur just before the maxima of the field oscillation, with the density at $QP$ energy reaching its local maxima.
 
\begin{figure}[h!]
\includegraphics[width=0.9\linewidth,angle=0]{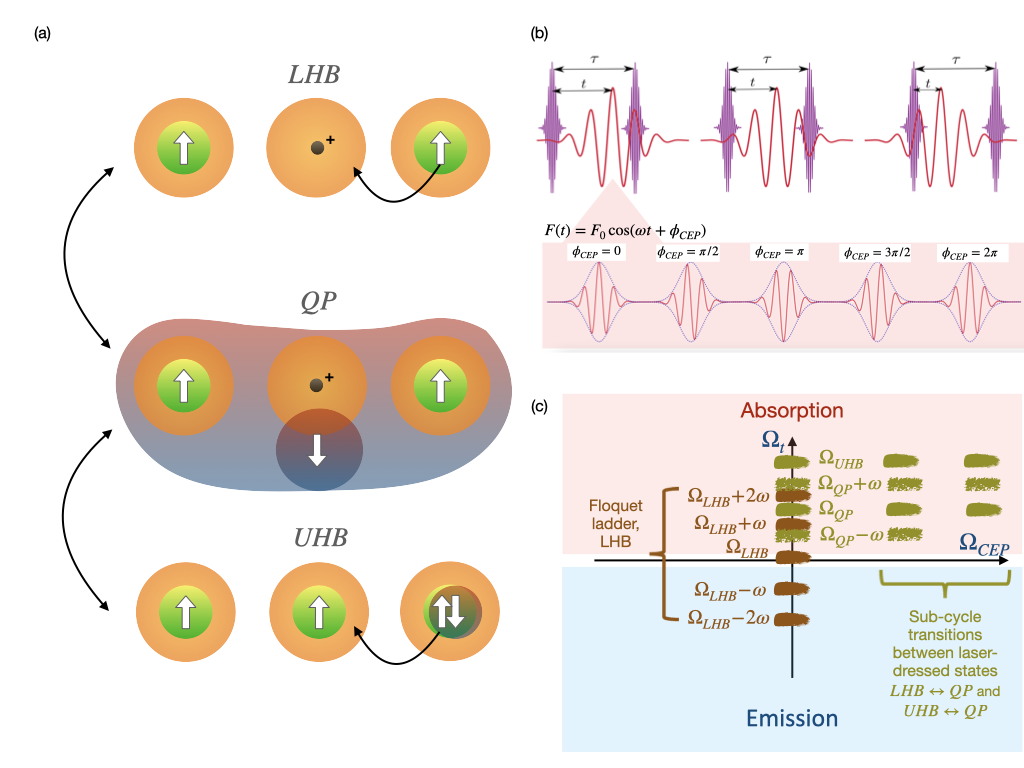}\caption{Spectroscopic nature of the Keldysh Green function $G^{<}$. (a) Cartoon view of the key many-body states  corresponding to the spectroscopic signal at the energies of $LHB$, $QP$, $UHB$. Orange circles stand for lattice sites, green circle symbolises electron localized on a  lattice site, white arrows show orientation of electron spin, 
 gray circle and a grey cloud symbolize delocolized electron, two oppositely oriented white arrows symbolize a doubly occupied site. Double headed arrows indicate possible sub-cycle transitions driven by the field in the phase locking regime (see Fig. 1 (f)). (b) Scanning delay $\tau$ between the pump and probe  pulses (violet), and delay $t$ between the pump-probe pair and the control pulse (red), yields two dimensional data set for the Green function $G^{<}(t,t-\tau)$ emulating photionization signal. Scanning the carrier-envelope phase ($\phi_{CEP}$) of the control pulse yields the third dimension of the spectroscopic signal $G^{<}(t,t-\tau,t_{CEP})$.  (c) Fourier transform of $G^{<}(t,t-\tau,t_{CEP})$ with respect to all arguments yields  $G^{<}(\Omega_{\tau},\Omega_{t},\Omega_{CEP})$. Cartoon view of $|G^{<}(\Omega_{\tau},\Omega_{t},\Omega_{CEP})|$ for  $\Omega_{\tau}$ fixed at the energy corresponding to $LHB$. Red-brown peaks illustrate the Floquet ladder associated representing laser-dressed $LHB$ state. The appearance of green peaks at $QP\pm n\omega$ and $UHB$ is due to non-adiabatic transitions between the laser dressed states $LHB\leftrightarrow QP$ and $UHB\leftrightarrow QP$. The extension of green peaks in the $\Omega_{CEP}$ dimension quantifies the sub-cycle response time. }
\label{Fig2}  
\end{figure}

The locking of populations at three key energies of the system 
%is a remarkable effect, which appears to be 
is correlated with the onset of metal to insulator transition observed in Figs. \ref{Fig1}(c,e). Cartoon in Fig. \ref{Fig2}(a) illustrates the three key field-free many-body states  of our system using the language of DMFT \cite{kotliar2004strongly}. Characteristic many-body states contributing to the signal at $LHB$ mainly involve   electrons localized on singly occupied lattice sites, $QP$ represents superposition of delocalized and localized electrons, while doubly occupied and unoccupied sites are the characteristic features of many-body states contributing to $UHB$.
Analysis (see SI) of Fig. \ref{Fig1} (f) suggests that the rate of flow of electron density \textit{from} $LHB$ and $UHB$ bands maximizes near zeroes of field oscillation ($F(t)\simeq 0$), while the rate of flow of electron density \textit{to} $LHB$ and $UHB$ bands maximizes near instantaneous maxima 
of the field ($|F(t)|\simeq F_0$).   %This flow is indicated by the double headed arrows in Fig. \ref{Fig1} (f). 
The oscillations of the density at $QP$ are in phase with the 
laser field: the minima coincide with $F(t)\simeq 0$, the maxima coincide with $|F(t)|\simeq F_0$. 

To decode the  underlying physics
%the laser driven phase transition in our system 
and  establish the role of sub-cycle electron dynamics we need a spectroscopy sensitive to such dynamics, which we introduce below. To this end, we exploit the full spectroscopic nature of the one particle  Keldysh Green's function $G^{<}(t,t-\tau)$, which can be retrieved from multi-pulse time and angular-resolved photoemission (trARPES) experiments (see e.g \cite{randi2017bypassing} and SI.) 
%, and conventional trARPES in solids does not use the full spectroscopic information in $G^<$. 

Figures \ref{Fig1}(b-d) show the Fourier image of 
such signal with respect to $\tau$, yielding the population of the 
occupied states (with energy $\Omega_{\tau}$)  prior to photo-ionization. The  
intense low-frequency field   plays the role of a control pulse, 
which modifies our system between these two events.
Thus, formally, we have a sequence of  three pulses (Fig.\ref{Fig2}(b)) reminiscent, but not identical, to the set-up of non-linear 2D spectroscopy \cite{reimann2021two-dimensional}.  Fourier transforming $G^{<}(t,t-\tau)$  with respect to $\tau$, selecting a value of $\Omega_{\tau}$, and Fourier transforming $G^{<}(t,\Omega_{\tau})$ with respect to the delay $t$ between the control pulse and the pump-probe pair, we obtain the spectrum of the states (transition frequencies)
%excited by the control pulse from $\Omega_{\tau}$, i.e. the frequencies $\Omega_{t}$ of the laser-dressed system corresponding to the Floquet quasienergies (see Methods). Moreover, Fourier transforming the $t$-dependent dynamics at fixed $\Omega_{\tau}$, we also get access to the states 
that the Floquet state associated with $\Omega_{\tau}$ couples to (see Methods). Indeed, for the dynamics described by a single Floquet state, $G^{<}(t,\Omega_{\tau})$ 
should behave periodically as a function of $t$, and its Fourier transform will
only show the sidebands at $\pm n\omega$ (n is integer, see the ladder of red-brown peaks in Fig.\ref{Fig2}(c)).
In contrast, in the presence of non-adiabatic transitions between the Floquet states, ${G}^{<}(t,\Omega_{\tau})$ becomes aperiodic and its Fourier transform will show the new frequencies appearing due to non-adiabatic excitations 
(see green peaks in Fig.\ref{Fig2}(c).)  

Next, we can reveal  the underlying sub-cycle dynamics by scanning the CEP $\phi_{CEP}$ of the IR (control) pulse and recording the resulting response $G^{<}(t,t-\tau,t_{CEP})$ as a function of $t_{CEP}=\phi_{CEP}/\omega$. The Fourier transform of $G^{<}(t,t-\tau,t_{CEP})$ with respect to $t_{CEP}$ shows the speed of response\cite{PhysRevLett.99.220406}: from instantaneous to cycle averaged. 
 The broader the resulting spectrum is with respect to the CEP 
(see schematic in Fig.\ref{Fig2}(c)), 
the stronger is the non-trivial 
CEP dependence, the stronger, faster, and more sensitive to the instantaneous electric field are the non-adiabatic 
transitions between the Floquet states. 
%%%%%%%%%%
We now focus on  non-adiabatic dynamics to highlight the part of the interaction, which is fundamentally different from the cycle-averaged response. To this end, we consider the difference between the derivatives of the Green's function with respect to $t$ and  $t_{CEP}$, $\Delta G^{<}_{ij}(t,t-\tau,t_{CEP})$, ($\Delta \equiv \frac{\partial }{\partial{t}}-\frac{\partial }{\partial{t_{CEP}}}$). $\Delta G^{<}_{ij}(t,t-\tau,t_{CEP})$ contains only non-adiabatic transitions
and provides direct access to non-equilibrium two-particle dynamics via the respective Green's functions  $K_{ij\sigma}^{pm}(t,t-\tau)$ and $\tilde{K}_{ij\sigma}^{pm}(t,t-\tau)$ (Methods):
\begin{eqnarray}
\label{Delta_G1}
\Delta G^{<}_{ij\sigma}=-\sum_{m,p} f(t-\tau)a_{mp}(t-\tau)K_{ij\sigma}^{pm}(t,t-\tau)- \sum_{m,p} f(t)a_{mp}(t)\tilde{K}_{ij\sigma}^{pm}(t,t-\tau)+\Delta G^{(1)<}_{ij\sigma},
\label{Delta_G2}
\end{eqnarray}
where $f(t)$ is the envelope of the short pulse, $a_{mp}(t)=\langle\Psi_m(t) |\frac{\partial }{\partial{f}}|\Psi_p(t)\rangle$ is an amplitude of non-adiabatic transition between time-dependent states evolving from the field-free eigenstates $|\Psi_m(t_0) \rangle$,  $|\Psi_p(t_0) \rangle$, $\Delta G^{(1)<}_{ij}$ are the non-adiabatic terms of one particle nature, and 
\begin{eqnarray}
\label{K_nonequilib1}
  K_{ij\sigma}^{pm}(t,t-\tau)=U\langle \Psi_p(t_0)|  c^{\dagger}_{i\sigma}(t-\tau)n_{i\overline{\sigma}}(t-\tau) c_{j,\sigma}(t)|\Psi_m(t_0) \rangle,\\
   \tilde{K}_{ij\sigma}^{pm}(t,t-\tau)=U\langle \Psi_p(t_0)|  c_{j\sigma}(t)n_{j\overline{\sigma}}(t) c^{\dagger}_{i,\sigma}(t-\tau)|\Psi_m(t_0) \rangle.
   \label{K_nonequilib2}
\end{eqnarray}
%\begin{equation}
%\label{K_nonequilib}
% K_{ij\sigma}^{pm}(t,t-\tau)=U\langle \Psi_p(t_0)|  c^{\dagger}_{i\sigma}(t-\tau)n_{i\overline{\sigma}}(t-\tau) c_{j,\sigma}(t)|\Psi_m(t_0) \rangle.
%\end{equation}
Here $U$ is the on-site Coulomb interaction,
$c_{i\sigma}^{\dagger}$ ($c_{j\sigma}$) are the fermionic creation (annihilation) operators for site $i$ ($j$) and spin $\sigma$, 
$n_{i\sigma}=c_{i\sigma}^{\dagger}c_{i\sigma}$ is the 
particle number operator.

%, evaluated at times separated by the pump-probe delay $\tau$. 
%5First, the time derivative of the one particle Green's function represented by the first term in Eq.(\ref{magic_formula}) is known to contain the two-particle Green's function.
%Second, we show that the second term, the derivative of the one particle Green's function with respect to $t_{CEP}$ (i) contains three-particle Green's function and (ii) cancels out the major part of one particle response lacking the CEP dependence from the first term, leaving us with only sub-cycle one-particle terms and all (sub-cycle and cycle-averaged) many-body contributions (see Methods).
% At the same time, time derivative of the Green's function  leaving the many
%This result shows unique power of sub-cycle response to record and address many-body dynamics in the most efficient way.

%Individual Floquet states lack CEP dependence, apart from the trivial dependence on $t-t_{CEP}$, which we explicitly remove upon the Fourier transform (see Methods).

%is the system to the instantaneous electric field,  the faster is its response, the stronger the non-adiabatic transitions between the Floquet states (the latter lack CEP dependence apart from the trivial dependence on $t-t_{CEP}$).

\begin{figure}[h!]
\begin{minipage}[h]{0.49\linewidth}
\begin{overpic}[width=1\textwidth]{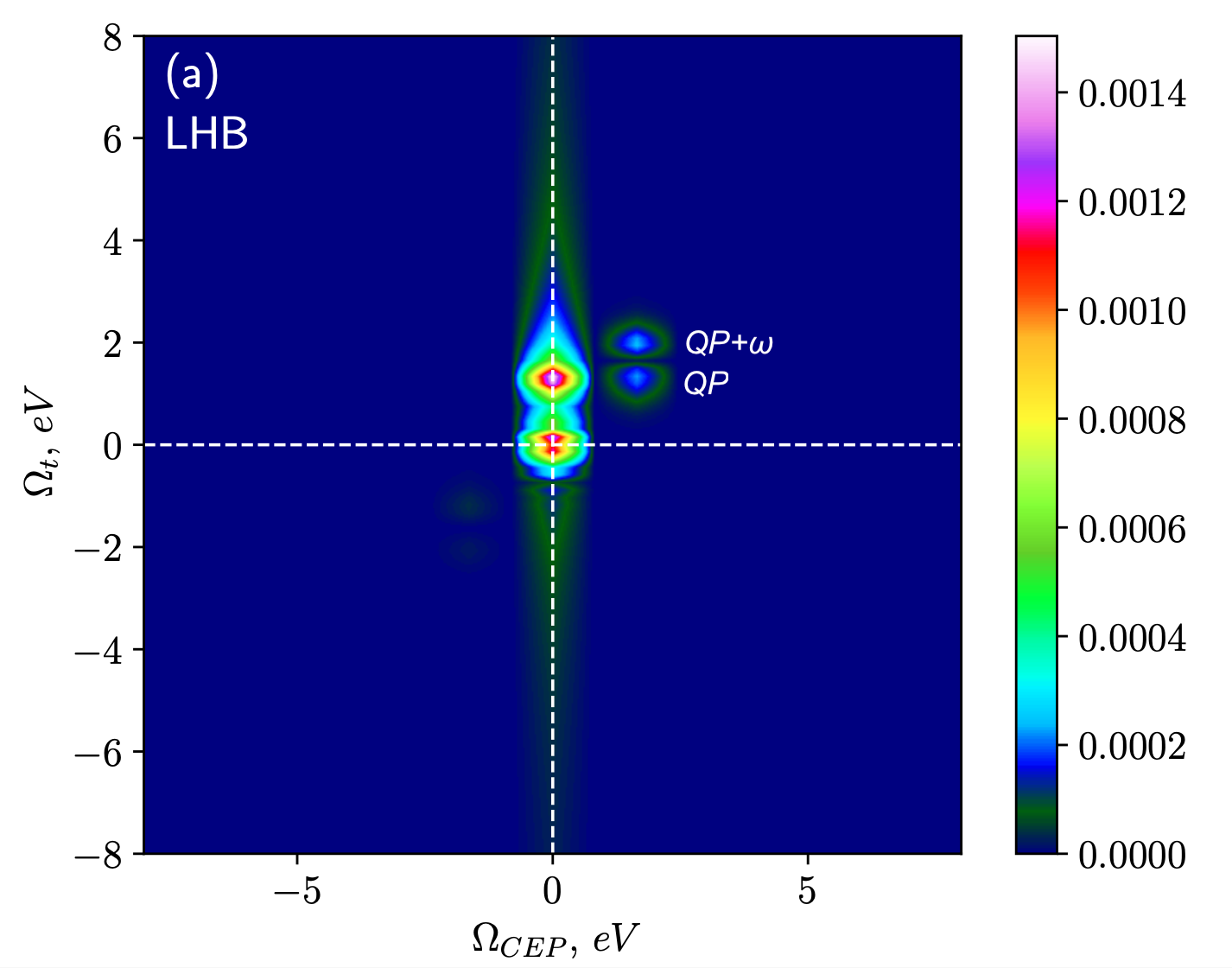}
 %\put (14.6,50.5) {\textcolor{white}{(a)}}
\end{overpic}
\end{minipage}
\hfill
\begin{minipage}[h]{0.49\linewidth}
\begin{overpic}[width=1\textwidth]{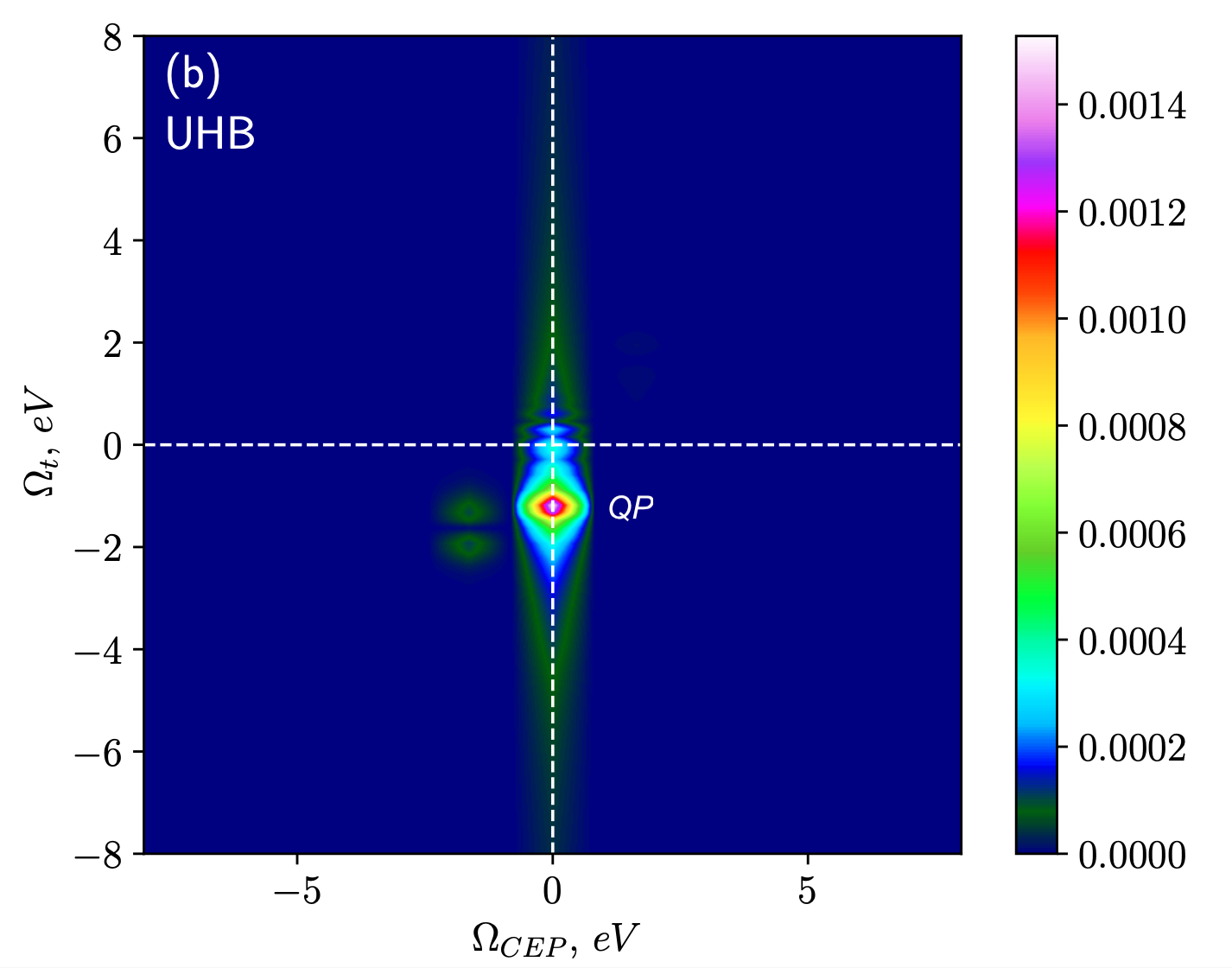}
% \put (21,58) {\textcolor{white}{(b)}}
\end{overpic}
\end{minipage}
\begin{minipage}[h]{0.49\linewidth}
\begin{overpic}[width=1\textwidth]{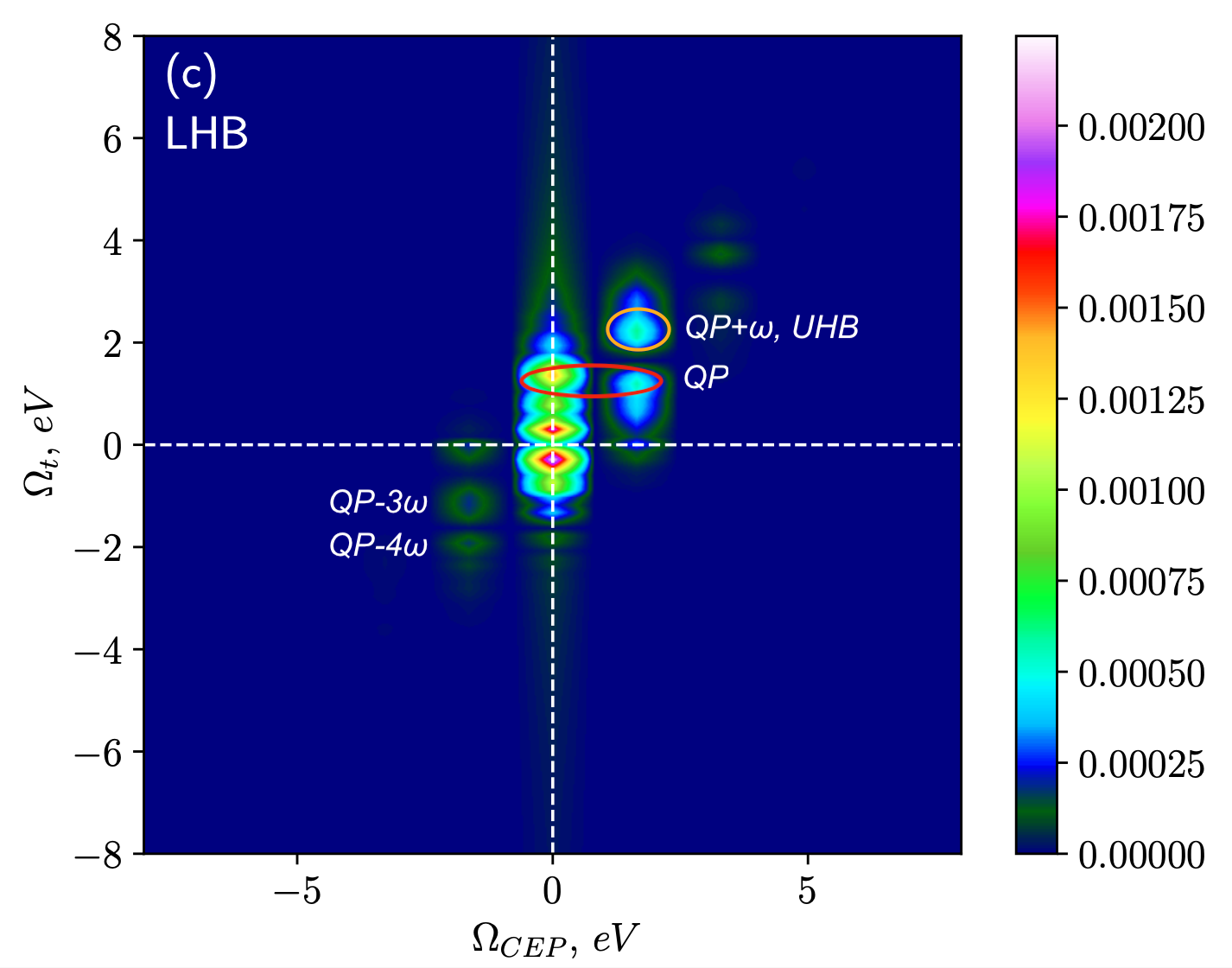}
% \put (14.6,50.5) {\textcolor{white}{(c)}}
\end{overpic}
\end{minipage}
\hfill
\begin{minipage}[h]{0.49\linewidth}
\begin{overpic}[width=1\textwidth]{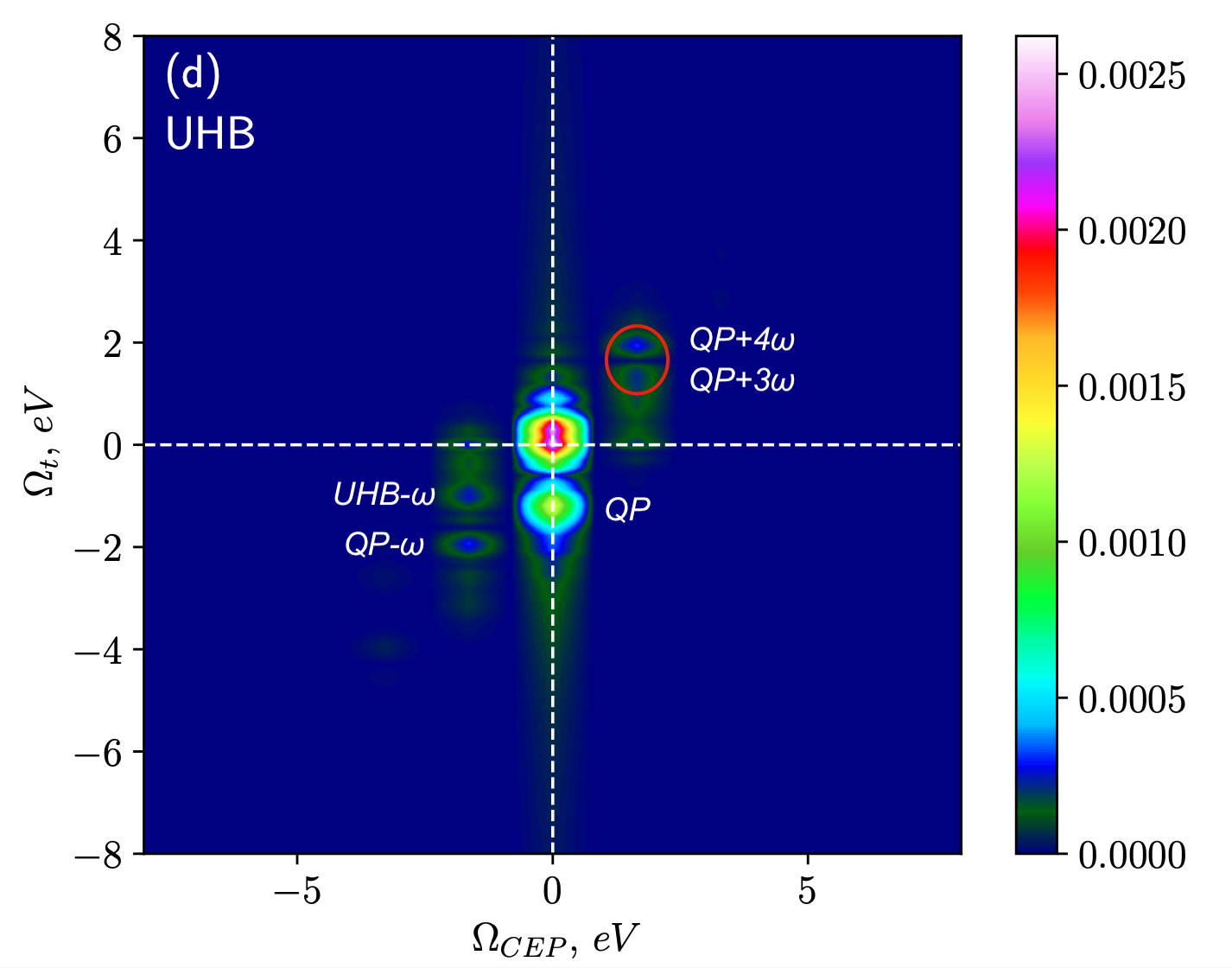}
 %\put (21,58) {\textcolor{white}{(d)}}
\end{overpic}
\end{minipage}
\begin{minipage}[h]{0.49\linewidth}
\begin{overpic}[width=1\textwidth]{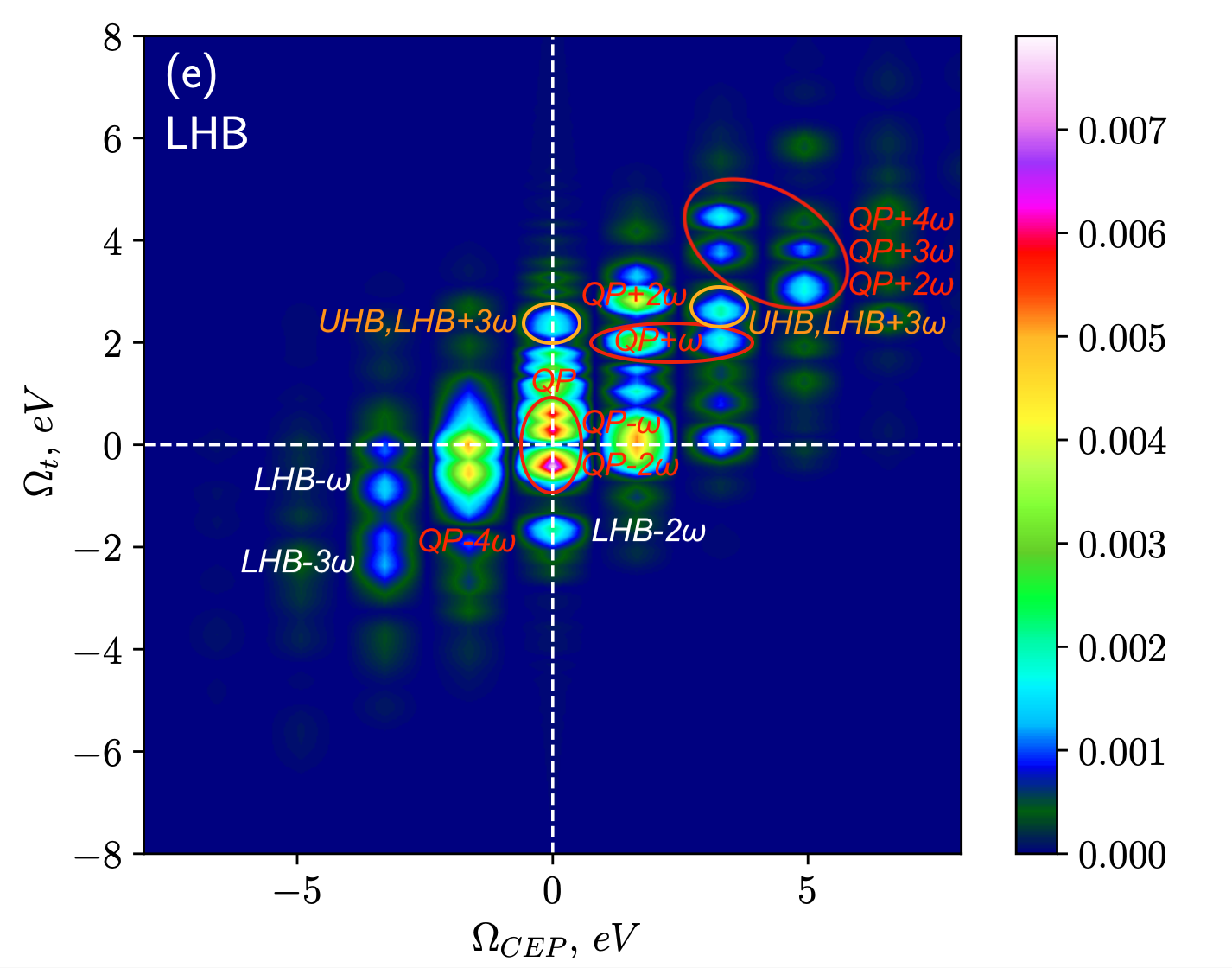}
% \put (14.6,50.5) {\textcolor{white}{(e)}}
\end{overpic}
\end{minipage}
\hfill
\begin{minipage}[h]{0.49\linewidth}
\begin{overpic}[width=1\textwidth]{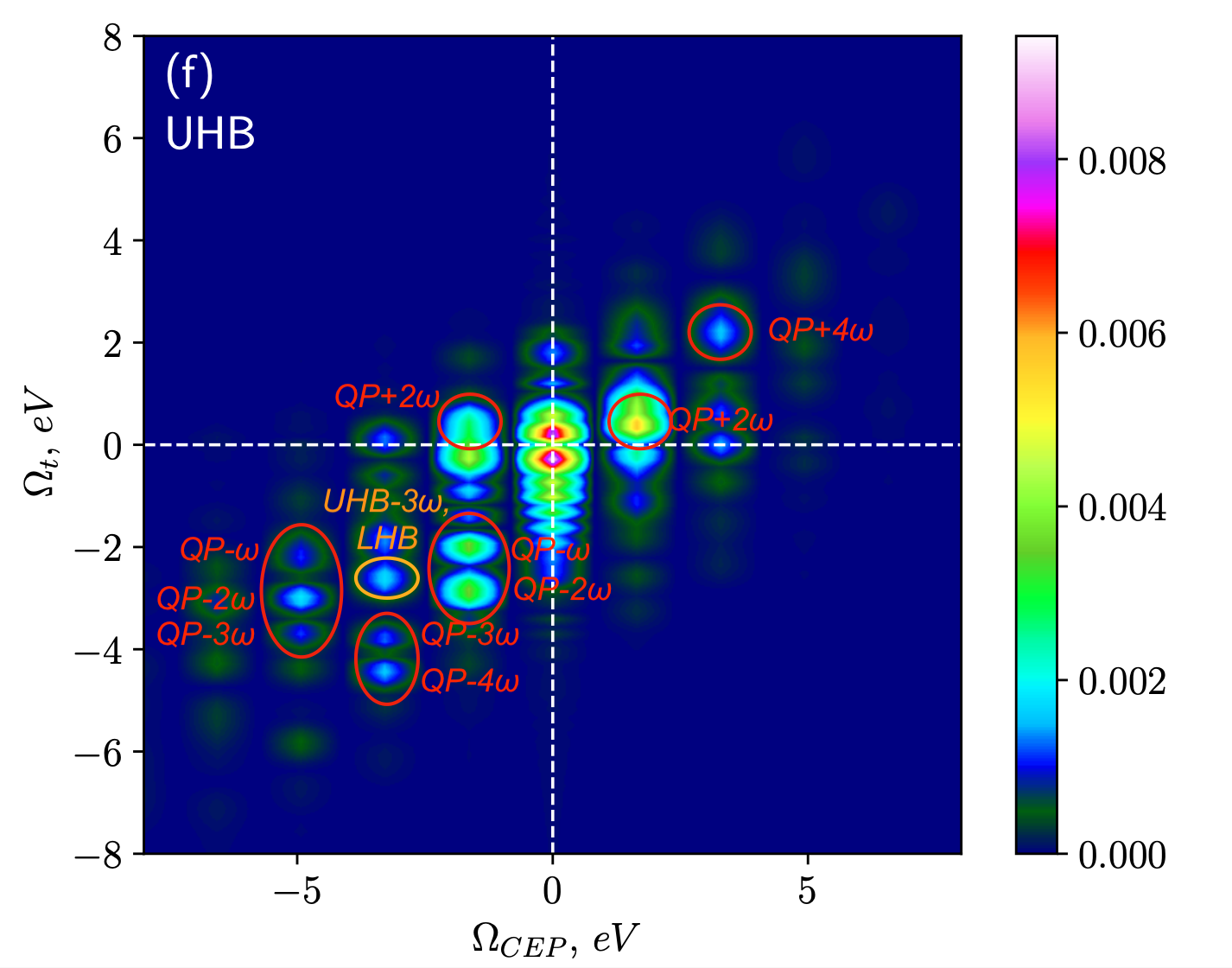}
 %\put (21,58) {\textcolor{white}{(f)}}
\end{overpic}
\end{minipage}
\caption{Multidimensional spectroscopy of correlated electron dynamics.
$|\Delta G^{<}(\Omega_{\tau},\Omega_{t},\Omega_{CEP})|$ vs $\Omega_{t},\Omega_{CEP}$ for  $\Omega_{\tau}=-1.25$ eV (LHB) (a,c,e) and $\Omega_{\tau}=1.25$ eV (UHB) (b,d,f) and different field strengths (a,b)  $F_0=0.1$ V/A, (c,d)  $F_0=0.5$ V/A, and (e,f) $F_0=2$ V/A.}
\label{Fig3}
\end{figure}

To visualize the non-adiabatic transitions encoded in $\Delta G^{<}(\Omega_{\tau},\Omega_{t},\Omega_{CEP})$  we fix $\Omega_{\tau}$ and plot the resulting 2D spectrum $|\Delta G^{<}(\Omega_{\tau},\Omega_{t},\Omega_{CEP})|$ as a function of $\Omega_{t}$ (vertical axis) and $\Omega_{CEP}$ (horizontal axis), as schematically shown in Fig.\ref{Fig2}(c). 

%First, for each $\Omega_{\tau}$, the peaks appearing at  $\Omega_{t}=0$ correspond to energy specified by $\Omega_{\tau}$. %i.e. represent diagonal terms $\Omega_{t}=\Omega_{\tau}$ in our frequency matrix. 
%Second, $\Omega_{t}>0$ correspond to absorption from $\Omega_{\tau}$, $\Omega_{t}<0$ correspond to emission from $\Omega_{\tau}$.   Third, peaks at $\Omega_{CEP}=0$ correspond to cycle-averaged (or one-photon) transitions.  The spectral peaks along the CEP dimension are separated by even number of photons due to the left-right symmetry of infinite lattice. Their strength reflects the importance of sub-cycle dynamics and the ability of the system to respond to the instantaneous field.
Figure \ref{Fig3}  shows the respective 2D spectra  $\Delta G^{<}(\Omega_{\tau},\Omega_{t},\Omega_{CEP})$ for  three intensities and for  $\Omega_{\tau}=-1.25$ eV (LHB),  $\Omega_{\tau}=1.25$ eV (UHB).
At the lowest field $F_0=0.1 V/A$ and for $\Omega_{\tau}=-1.25$ eV, the two peaks  dominating the spectrum Fig.\ref{Fig2}(a) at $\Omega_{CEP}=0$ correspond to $LHB$ ($\Omega_{t}=0$) and $QP$ ($\Omega_{t}\simeq 1.25$ eV). % The coupling between $LHB$ and $QP$ is already quite strong: indeed, $F_0a_0\sim E_{QP}-(\omega+E_{LHB})\sim \omega$ already at this field strength. 
%Since the field-free system starts in $QP$, the $QP\leftrightarrow LHB$ coupling can be interpreted as localization due to field-induced suppression of tunnelling on the sub-cycle scale, as evidenced by the additional peaks at $\Omega_{CEP}=2\omega$.
Figure \ref{Fig3}(b) shows the 2D spectra for $\Omega_{\tau}=1.25$ eV corresponding to $UHB$, for the same $F_0$. The dominant peak at $\Omega_{CEP}= 0$ corresponds to the contribution of $QP$, signifying transitions between  $QP$ and $UHB$. %The relative weakness of peaks at $\Omega_{CEP}\neq 0$ shows that their coupling is well described by cycle-averaged processes. 
\begin{figure}[h!]
\begin{minipage}[h]{0.49\linewidth}
\begin{overpic}[width=1\textwidth]{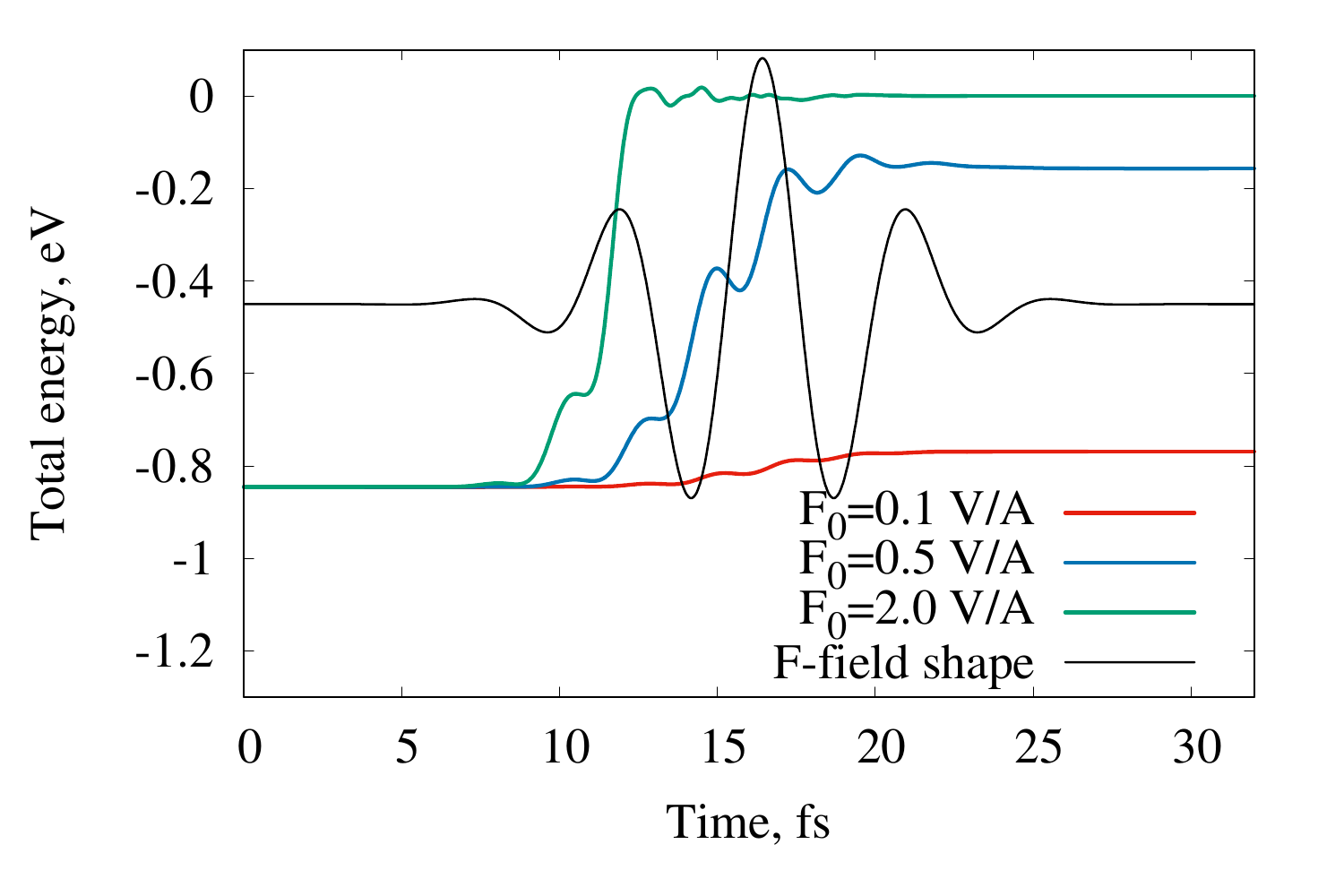}
 \put (20,58) {\textcolor{black}{(a)}}
\end{overpic}
\end{minipage}
\hfill
\begin{minipage}[h]{0.49\linewidth}
\begin{overpic}[width=1\textwidth]{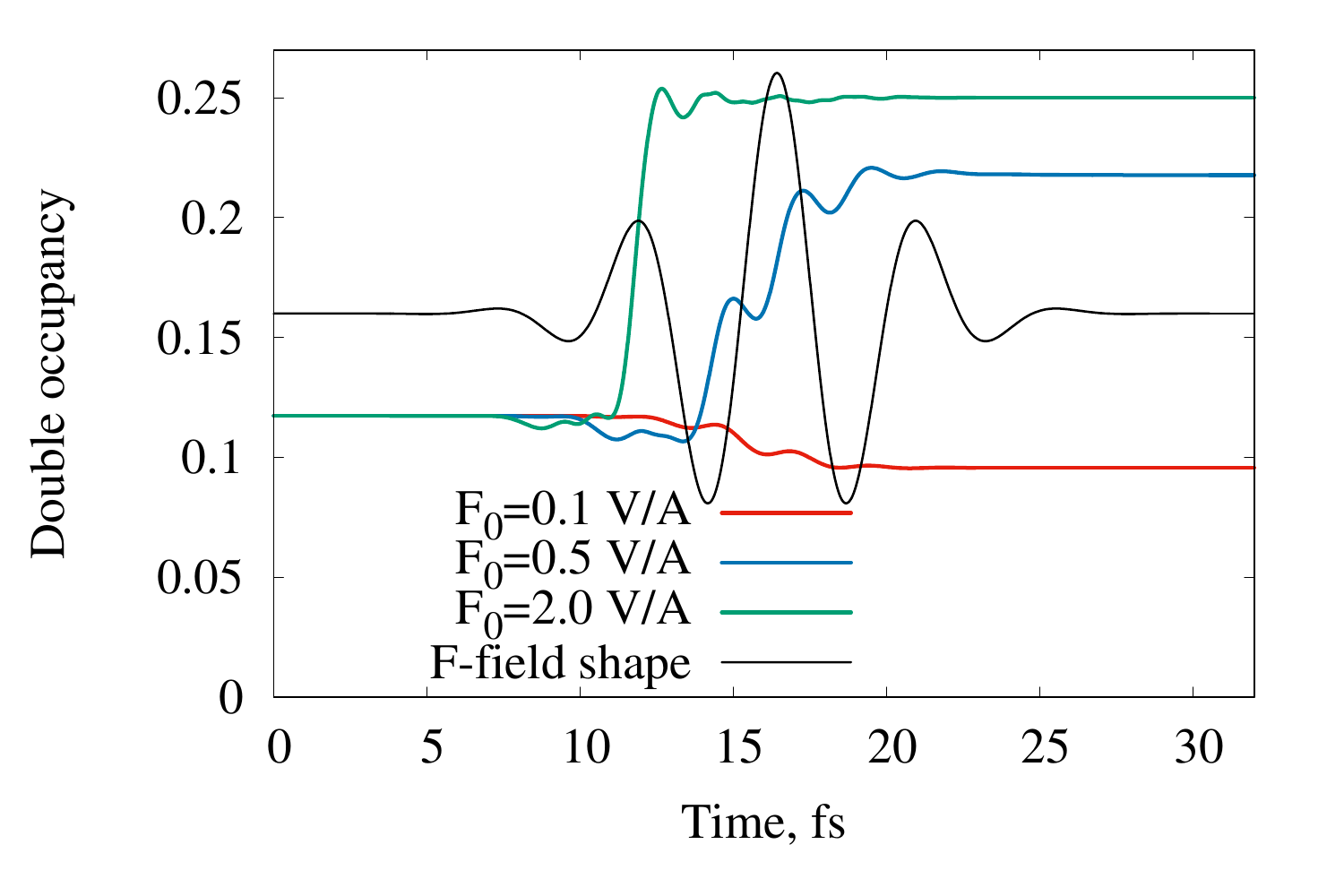}
 \put (22.5,58) {\textcolor{black}{(b)}}
\end{overpic}
\end{minipage}
\begin{minipage}[h]{0.49\linewidth}
\begin{overpic}[width=1\textwidth]{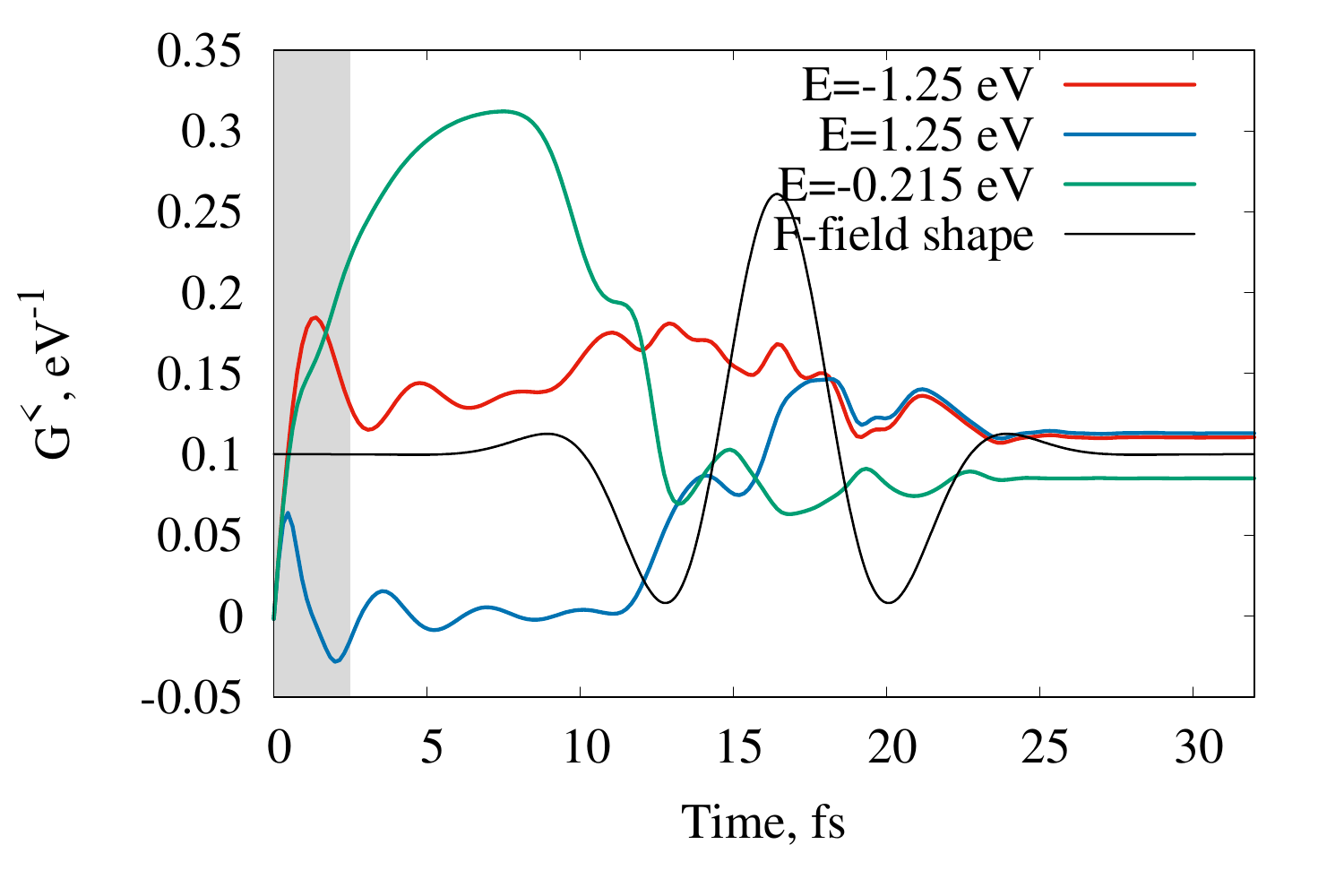}
 %\put (14.5,62) {\textcolor{black}{(c)}}
 \put (21.5,58) {\textcolor{black}{(c)}}
 %\put (14.5,33) {\textcolor{white}{(d)}}
\end{overpic}
\end{minipage}
\hfill
\begin{minipage}[h]{0.49\linewidth}
\begin{overpic}[width=1\textwidth]{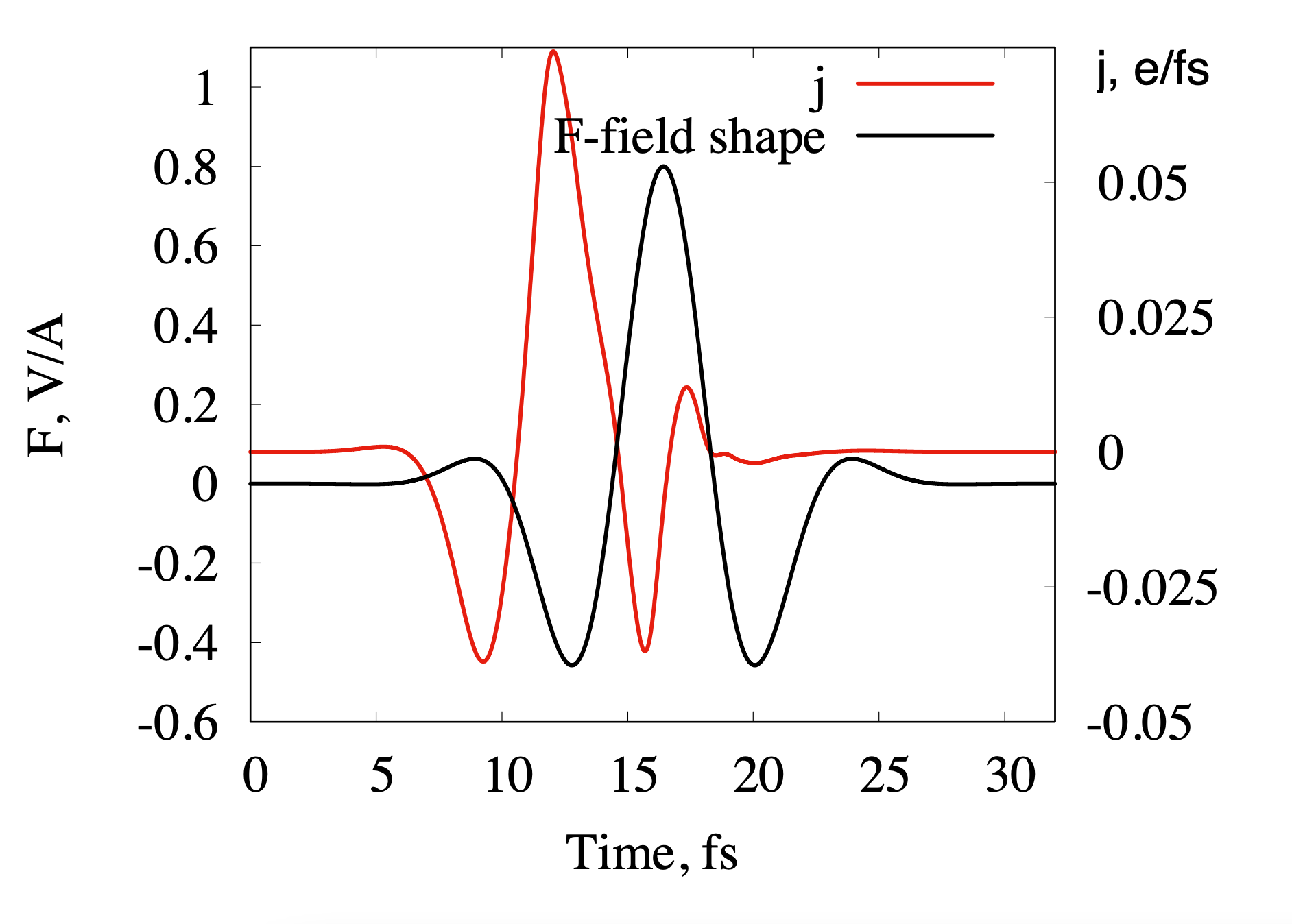}
 \put (21.5,63.5) {\textcolor{black}{(d)}}
\end{overpic}
\end{minipage}
\caption{Temporal dynamics of metal-insulator transition. (a,b) Time-resolved total energy (a) and doublon occupancy (b) for $\omega=0.827$ eV and various field strengths:  $F_0$=0.1 V/A (red),  0.5 V/A,(blue) 2.0 V/A (green). (c,d) Metal-insulator transition driven by the laser field carried at $\omega=0.413$ eV. (c) Oscillations  of electron density at energies corresponding to $LHB$  ($E=-1.25$ eV, red), $UHB$ ($E=1.25$ eV, blue), and $QP$ ($E=-0.215$ eV, green) for $F_0=0.8$ V/A. (d) Laser pulse (black) and current (red).
Onset of locking regime at ~18 fs (c) coincides with quenching of current (d).}
\label{Fig4}
\end{figure}

At the intermediate field $F_0=0.5$ V/A, (Fig \ref{Fig3}(c,d)) the peaks at $\Omega_{CEP}= 0$ in Fig. \ref{Fig3}(c) (for  $\Omega_{\tau}=-1.25$ eV, LHB) correspond to the  Floquet ladders associated with both QP and LHB, with broad overlapping steps. The CEP-sensitive dynamics becomes more significant in the LHB--QP transitions. 
The dominant part of the upper peak at $\Omega_{CEP}=2\omega$ is due to the $QP$ lifted by one photon or $UHB$.

The spectroscopic portrait of the system in the high-field  regime  ($F_0=2$ V/A), Fig \ref{Fig3}(e,f) contains several new features. First, we see broad spectrum along the vertical axis ($\Omega_{t}$) for a wide range of horizontal axis ($\Omega_{CEP}$). Second, the sub-cycle dynamics is very important and the spectrum has individual sub-cycle cut-offs: the highest  positive and highest negative $\Omega_{t}$ depend on $\Omega_{CEP}$.
%Check: It probably  fits into sub-cycle cut-off expected in the strongly driven two level system with the gap $\Delta E$: ($\Omega_{t}^{cut-off}=\sqrt{2|F_{eff}\cdot a|^{2}+|\Delta E|^2}$).
These cut-offs appear to be proportional to the instantaneous values of the laser field  and increase with increasing $\Omega_{CEP}>0$. Third, 
the overall 2D spectra are shifted towards positive values of $\Omega_{t}$ for LHB, corresponding to absorption, and negative values of $\Omega_{t}$ for UHB, corresponding to emission.
The respective peaks at $\Omega_{CEP}=4\omega$ (Fig \ref{Fig3}(e)) and $\Omega_{CEP}=-4\omega$ (Fig \ref{Fig3}(f)) indicate their strongly sub-cycle nature.
Forth, while the direct non-adiabatic transitions between LHB and UHB (orange circles in Fig \ref{Fig3}(e)) become visible, the 
non-adiabatic transitions are still dominated by the  couplings $LHB\leftrightarrow QP$ and $UHB\leftrightarrow QP$.   

Indeed, the onset of locking is synchronized  with the  saturation of energy transfer from the field to the system 
%-- the instantaneous total energy of the system during the pulse becomes frozen at its maximum (for a given field strength) value (
(see Fig \ref{Fig4}(a)). For the highest field ($F_0=2 V/A$) the maximum energy saturates at zero.   %The maximum fraction of doublons is $0.25$, corresponding to the uncorrelated limit (see Fig \ref{Fig4}(b)). This observation suggests that the locking regime effectively corresponds to the strongly driven single particle dynamics. %This conclusion is further supported by the 2D spectra in high field  regime  ($2 V/A$)(see Fig \ref{Fig2}}(e,f)), where the "locking" phase of electron dynamics dominates the signal (see Fig \ref{Fig2}(c)}).
The onset of locking (see Fig \ref{Fig4}(c)) is accompanied by the suppression of the current (see Fig \ref{Fig4}(d)), which is fully quenched 
at $\sim 18$ fs (see Fig \ref{Fig4}(c,e)) when  the insulating state is established. Note that %the results presented in 
Figs \ref{Fig4}(c,d) correspond to the lower laser frequency $\omega=0.413$ eV, demonstrating that our results are  not sensitive to the frequency of the non-resonant field.

%Overall, the onset of locking regime is associated with very broad distribution of overlapping laser-dressed states  (see Fig \ref{Fig3}(e,f)), effectively forming a single band available for electron dynamics. Synchronized cycling of charge flow along the two "circuits"  -- $LHB\rightarrow QP \rightarrow LHB$ and $UHB\rightarrow QP \rightarrow UHB$ -- appear to dominate the dynamical laser-dressed many-body state just before it freezes into the final state after the end of the laser pulse.

This spectroscopic information points to the following simplified  picture of  many-body dynamics in the  locking regime.  The synchronized cycling of charge flow along the two "circuits," $LHB\rightarrow QP \rightarrow LHB$ and $UHB\rightarrow QP \rightarrow UHB$,  appears to dominate the dynamical photon-dressed many-body state just before it freezes into the final state after the end of the laser pulse. At the peak of the field, a localized electron in LHB is promoted to the QP, leaving an empty site behind. At the same time, the strong field also destroys doublons: a doublon from UHB loses one of its electrons into the QP. The total energy does not change during these two synchronized processes. Near the instantaneous zero of the field, the QP electrons  localize  on the empty or half-filled sites of the lattice completing the oscillation cycle:  localization at an empty site contributes to the population in LHB, while localization  at a half-filled site contributes to the population in UHB. However, in contrast to light-driven dynamics in dielectrics\cite{vampa2015linking}, the flow of charge between   LHB and QP or  UHB and QP cannot be understood as single electron oscillation.

Indeed, the time-evolution of double occupancy (Fig.\ref{Fig4}(b)) shows that in the locking regime the  doublon production saturates at $0.25$, the maximal value for an uncorrelated system. Thus, every produced doublon decays and its fragments randomly occupy lattice sites.  This "randomness"  limits the total energy increase to zero energy and  precludes population inversion, i.e. achieving higher electron density at $UHB$ compared to $LHB$. This apparent randomness could be a sign of  entanglement destroyed by our observation. For example, calculating double occupancy we trace out a multitude of different ways in which  doublons are crated and destroyed each laser cycle, i.e. 
we do not follow their individual Feynman paths when calculating this observable and therefore we partially destroy the entanglment in our system.

In correlated systems, large increase in electron temperature can transform a metal into a bad metallic or insulating-like state. 
%ME
However, in our case, 
%the state thermalizes into such a high-temperature bad metallic state few femtoseconds within few inverse hoppings after the pulse, 
the opening of the gap and the peculiar  dynamics observed already within a small fraction of the laser cycle is clearly not thermal.
%However, in our case the transformation is not thermal -- it occurs within a small fraction of the laser cycle, on the time-scale too short for thermalization, although thermalization may happen at later times after the end of the pulse. 
In contrast to phonon-driven transitions \cite{forst2011nonlinear}, 
our mechanism is purely electronic.

%along with single particle dynamics $S_{ij}$ (see Methods) and fully correlated non-adiabatic transitions  $\Delta G^{<}_{ij}(t,t-\tau,t_{CEP})$.
Our results show the power of sub-cycle response to record and address non-equilibrium many-body dynamics. Here we considered the first derivative with respect to $\Delta \equiv \frac{\partial }{\partial{t}}-\frac{\partial }{\partial{t_{CEP}}}$, which gave us access to non-equilibrium two-particle correlations.  Likewise, the $n$-th  derivative, $\Delta^n$, contains  the non-equilibrium $n$-particle Green's functions. If the correlations are strong, as is the case here,  odd  and even multiphoton pathways contributing to the CEP-dependent response can involve photon absorption by different electrons, with correlated interaction establishing  coherence between these interfering events. Note that in contrast to the standard equilibrium (diagonal in m)  expression for the two-particle Green's functions $K_{ij\sigma}^{mm}(t,t-\tau)$ and $K_{ij\sigma}^{mm}(t,t-\tau)$,
$\Delta G^{<}_{ij}$ features non-equilibrium two-particle correlations, which would not be recorded in long pulses, as $\Delta G^{<}_{ij}= 0$ in this case, corresponding to the standard  Floquet regime. 
%Such step by step recovery of many body Greens functions is explicit in long pulse two-color scenario, where the relative phase between the two colors $\phi_{\omega,2\omega}$ plays a role of $t_{CEP}$, $t_{CEP}=\phi_{\omega,2\omega}/\omega$. In this case the fully correlated non-adiabatic term vanishes $\Delta G^{<}_{ij}(t,t-\tau,t_{CEP})$, but the array of many body Green's  functions can be directly recovered via derivatives of the signal with respect to the $\phi_{\omega,2\omega}$. }

%We show how many-body correlations introduce an
%important new aspect: the newly created states of the 
%laser-dressed system do not have to disappear as the 
%light is switched off. Altering the effective 
%potential for the electron motion with strong pulse light we not only
%convert the system from a metallic state to the state
%with a Mott gap, but also achieve its survival after the 
%light is turned off. 
% The dynamics in time-domain is here 
% analyzed via harmonic generation spectroscopy, which 
% encodes the formation of the Mott gap, 
% excitation dynamics across it, and the establishment 
% of the insulating state. 

Our findings demonstrate the possibility of  manipulating phases 
of correlated systems with strong non-resonant  
fields on the sub-cycle time-scale, in a manner that is robust against the 
frequency of the driving field. 
While in atoms or molecules such non-resonant light-induced modifications of the electron density vanish once the light is turned off, 
in a strongly correlated system light-induced changes can lead to persistent modifications surviving after the end of the pulse, as we see here.  
The pulse can thus transfer  the system into a correlated state inaccessible under equilibrium conditions, with non-adiabatic transitions triggering non-equilibrium many-body correlations, see Eq.(2).
%ME possibly: ... at least not when electrons and lattice are in equilibrium with each other.
%ME
%In particular, for systems close to the Mott transition or with multiple orbital degrees of freedom, this controlled sub-cycle dynamics can offer new mechanisms to control the many body phases, beyond the multi-cycle Floquet engineering.
%
The  achieved final state is 
controlled by charge density and currents shaped on the sub-laser-cycle time-scale;
%ME 
the spectroscopy introduced here can provide key insights in analysing and designing such excitation pathways. 
%These findings offer new mechanism of control  beyond the multi-cycle Floquet engineering.

%and remains more insulating than the initial state after the pulse.

%We demonstrate that one can generate new  properties of strongly correlated materials after the end of the
%pulse via sub-laser cycle modifications during the pulse and 

%Merlin.mbs v4.21 2009-07-09.

%\bibliography{reference}
% Produces the bibliography via BibTeX.

%\bibliography{correlatedsolids_}

\section*{Acknowledgements}
We thank M. Altarelli for motivating discussions, U. Bovensiepen and M. Ligges for discussions on pump-probe PES of cuprates,   
R. E. F. Silva for providing his code for benchmark and Y. Mohammed for checking the analytical calculations presented in Eqs (1-3) of the main text and useful suggestions. This research was supported in part through the European XFEL and DESY computational resources in the Maxwell infrastructure operated at Deutsches Elektronen-Synchrotron (DESY), Hamburg, Germany.  This work was supported by the Cluster of Excellence "Advanced Imaging of Matter" of the Deutsche Forschungs-gemeinschaft (DFG) - EXC 2056 - project ID390715994.
H.A. acknowledges a support from the ImPACT Program of
the Council for Science, Technology and Innovation, Cabinet
Office, Government of Japan (Grant No. 2015-PM12-05-01)
from JST, JSPS KAKENHI Grant No.17H06138, and CREST ``Topology" project from JST. 
M.I. and O.S. acknowledge support of the  H2020 European Research Council Optologic 
grant (899794). The work of A.I.L. and M.I.K. is supported by European Research Council via Synergy Grant 854843 - FASTCORR. M.E. acknowledges funding from the ERC via starting grant No. 716648. The work of Y. M. was funded by the European Union (ERC, ULISSES, 101054696). Views and opinions expressed are however those of the author(s) only and do not necessarily reflect those of the European Union or the European Research Council. Neither the European Union nor the granting authority can be held responsible for them. 
%Merlin.mbs v4.21 2009-07-09.

%\bibliography{reference}
% Produces the bibliography via BibTeX.

%\bibliography{correlatedsolids_}
\bibliography{arXiv_submission_10March}

\section{\label{Methods}Methods}
\subsection{\label{Simulations}Simulations}

The Hamiltonian is 
  \begin{align}
    \label{Ham1}
    H(t)
    &=
    \sum_{ij\sigma}
    T_{ij}(t)
    c_{i\sigma}^{\dagger}
    c_{j\sigma}
    +
    U
    \sum_{i}
    \big(n_{i\uparrow}\!-\!\tfrac12\big)
    \big(n_{i\downarrow}\!-\!\tfrac12\big)
    ,
  \end{align}
where $i, j$ label the lattice sites, 
%$T_{ij}(t)$ is hopping amplitude,  
$U$ is the on-site Coulomb interaction,
$c_{i\sigma}^{\dagger}$ ($c_{j\sigma}$) are the fermionic creation (annihilation) operators for site $i$ ($j$) and spin $\sigma$, 
$n_{i\sigma}=c_{i\sigma}^{\dagger}c_{i\sigma}$ is the 
particle number operator.
The hopping amplitudes $T_{ij}(t)$  between the 
sites  $i$ and $j$ include the 
nearest-neighbor ($T_1$) and the next-neighbor ($T_2$)
terms. The external low-frequency laser field 
(frequency $\omega < U, W=8T_1$) is included
via the Peierls substitution, 
  \begin{equation}
T_{ij}(t)=T_{ij}\exp{ \left( -i\int_{{\bf R}_{j}}^{{\bf R}_{i}} d{\bf r} \cdot {\bf A}(t) \right) },
\end{equation}
where ${\bf A}(t)$ is the field vector-potential,
${\bf F}(t) ={-} {\partial} {\bf A}(t) / {\partial t} $. The one-particle dispersion is
%\begin{equation}
%\varepsilon(k,t)={-2t(cos(k_x+{\bf A}_x(t))+cos(k_y+{\bf A}_y(t)))}\\
%\label{dispersion}
%\end{equation}
\begin{equation}
\begin{split}
  % {\varepsilon_k={2t_1(cos(k_x)+cos(k_y))+4t_2(cos(k_x)\cdot cos(k_y))+2t_3(cos(2k_x)+cos(2k_y))}}
 %  &{\varepsilon_1 (k,t)={2t_1(cos(k_x+{\bf A}_x(t))+cos(k_y+{\bf A}_y(t)))}};\\
 % &{\varepsilon_3 (k,t)={2t_1(cos(k_x+{\bf A}_x(t))+cos(k_y+{\bf A}_y(t)))+4t_2(cos(k_x+{\bf A}_x(t))\cdot cos(k_y+{\bf A}_y(t)))+}}\\
%  &{ { +2t_3(cos(2k_x+2{\bf A}_x(t))+cos(2k_y+2{\bf A}_y(t)))}}.
  % ek[k][tstp]=J1*2*(cos(kk(1)+A[tstp+1](1))+cos(kk(0)+A[tstp+1](0)))+J2*4*(cos(kk(1)+A[tstp+1](1))*cos(kk(0)+A[tstp+1](0)))+J3*2*(cos(2*(kk(1)+A[tstp+1](1)))+cos(2*(kk(0)+A[tstp+1](0))));
  \varepsilon(k,t)&=2T_1[\cos(k_x+{\bf A}_x(t))+\cos(k_y+{\bf A}_y(t))]\\
  &+4T_2[\cos(k_x+{\bf A}_x(t))\cdot \cos(k_y+{\bf A}_y(t))].
 %\\
 % &{ { +2t_3(cos(2k_x+2{\bf A}_x(t))+cos(2k_y+2{\bf A}_y(t)))}}.
\end{split}
\label{dispersion}
\end{equation}
% For the material-specific calculations of FS presented in Sec. \ref{piPulseFS} we use tight-binding model for $YBa_2Cu_3O_y$ (YBCO), adopted from Ref. \cite{ALYBCO}. We included only nearest neighbors and next nearest neighbors hoppings,  $T_1=0.69$~eV, $T_2=-0.22$~eV, since the essential features, e.g. FS topology are already captured by this model.
The total energy $E_\text{tot}(t) = E_\text{kin}(t) + E_\text{pot}(t)$ includes potential  and
 kinetic  terms, 
\begin{eqnarray}
 &&E_\text{pot}(t) = 
    U \big\langle \big(n_{\uparrow}\!-\!\tfrac12\big)
    \big(n_{\downarrow}\!-\!\tfrac12\big) \big\rangle, 
 % =  U ( d(t)-\tfrac12 \left[n_{\uparrow}+n_{\downarrow}\right] 
 %+\tfrac14 ),
  \label{Epot}
\\
 &&E_\text{kin}(t) = 
  -i \sum_{\bf{k}}{\varepsilon_{\bf{k}}{{G}_{{\bf k}+{{\bf A}(t)}}^{<}}(t, t)},
      \label{Ekin}
\end{eqnarray}
    % \begin{equation}
    %   E_\text{kin}(t) = 
    %   -i \sum_{\bf{k}}{\varepsilon_{\bf{k}}{{G}_{{\bf k}+{{\bf A}(t)}}^{<}}(t, t)},
    %   \label{Ekin}
    % \end{equation}
    % \begin{equation}
    %   E_\text{tot}(t) = E_\text{kin}(t) + E_\text{pot}(t)
    %     \label{Etot}
    % \end{equation}
    % respectively. Here $d(t)$ denotes time-dependent double occupancy of the correlated site.
% We calculate electron momentum distribution as Momentum distribution defined by $n=f({\bf k},t)=-i{{\tilde G}_{\bf k}^{<}(t,t)}=-i{{G}_{{\bf k}+{{\bf A}(t)}}^{<}(t,t)}$.
% % where ${{\tilde G}_{\bf k}^{<}(t,t)}$  $[{{G}_{\bf k}^{<}(t,t)}]$.
% The electric field $ E(t) ={-} {\partial} {\bf A}(t) / {\partial t} $, ${\bf A}(t)$ is vector potential.
where ${{\tilde G}_{\bf k}^{<}(t,t)}$  is the gauge-invariant \cite{tsuji2008correlated} lesser Green function. 
The momentum distribution function is
\begin{equation}
 n({\bf k},t)=f({\bf k},t)=-i{{\tilde G}_{\bf k}^{<}(t,t)}=-i{{G}_{{\bf k}+{{\bf A}(t)}}^{<}(t,t)}.
\label{Distr}
\end{equation}
The population density is calculated as
\begin{equation}
G^{<}(\omega,t) = \frac{1}{\pi}{\rm Im}\int ds e^{i\omega s}G^{<}(t,t-s).
\label{Gles}
\end{equation}
Due to the limitation in time data, the selected Fourier transform produces some blur on the graph of occupied states 
for the first 5 fs.

The time-resolved photoemission intensity is 
given by 
\begin{equation}
\label{PES}
I(\omega,t_p)=-i\int dtdt'S(t)S(t')e^{i\omega(t-t')}G^{<}(t+t_p,t'+t_p),
\end{equation}
where $S(t-t_p)$ is the envelope of the 
probe pulse centered at $t_p$ \cite{freericks2009theoretical}.

\subsection{\label{Non-adiabatic} Direct access to non-adiabatic many-body dynamics}
Here we develop the strategy for isolating the non-adiabatic response in the one particle Green's function.
In a long laser pulse, the dependence of the instantaneous electric field on the CEP amounts to
the overall time shift $t'=t+t_{CEP}$. 
This trivial dependence is of no interest and should be removed when analysing the Green's function $G^{<}_{ij}(t,t-\tau,t_{CEP})$. 
To see how this should be done, let us assume for the moment that its CEP dependence amounts only to the overall time shift of the argument $t'=t+t_{CEP}$: 
\begin{equation}
 \tilde G^{<}_{ij}(t,t-\tau,t_{CEP})=\tilde G^{<}_{ij}(t',t'-\tau).   
\end{equation}
If this is the case, the Green's function should obey the following equation:
\begin{equation}
 \frac{\partial  G^{<}_{ij}}{\partial t}(t,t-\tau,t_{CEP})= \frac{\partial G^{<}_{ij}}{\partial t_{CEP}}(t,t-\tau,t_{CEP}). 
\end{equation}
Therefore, differentiating the Green's function $G^{<}_{ij}(t,t-\tau,t_{CEP})$ with respect to $t$ and $t_{CEP}$ and subtracting the resulting terms we obtain the differential contribution $\Delta G^{<}_{ij}(t,t-\tau,t_{CEP})$, which no longer contains the trivial dependence:
\begin{equation}
\label{magic_formula_Methods}
 \Delta G^{<}_{ij}(t,t-\tau,t_{CEP})=\frac{\partial G^{<}_{ij}}{\partial t}(t,t-\tau,t_{CEP})-\frac{\partial G^{<}_{ij}}{\partial t_{CEP}}(t,t-\tau,t_{CEP}).
\end{equation}
We use $\Delta G^{<}_{ij}(t,t-\tau,t_{CEP})$ for building the 2D spectroscopy maps as a function of $\Omega_{t}$ and $\Omega_{CEP}$.

Now we can explicitly evaluate $\Delta G^{<}_{ij}(t,t-\tau,t_{CEP})$ (Eq.\ref{magic_formula_Methods} ) for an arbitrary  Hamiltonian $H(t)$. 

Since we consider coherent dynamics, we can rewrite Eq. (\ref{G_less}) for the Green's function as follows (see e.g. \cite{stefanucci_vanleeuwen_2013},\cite{kamenev_2011}):
\begin{equation}
 G^{<}_{ij}(t,t-\tau)= i \sum_m\langle \Psi_m(t_0)|c_{i}^{\dagger}(t-\tau)c_j(t)|\Psi_m(t_0)\rangle 
 \label{G_less}
\end{equation}
where $|\Psi_m(t_0)\rangle$ are field free eigenstates of the system, $t_0$ is the initial moment before the laser pulse. Inserting the resolution of identity $I=\sum_n |\Psi_n(t_0)\rangle\langle \Psi_n(t_0)|$ on the field free eigenstates
\begin{equation}
 G^{<}_{ij}(t,t-\tau)= i \sum_{m,n}\langle \Psi_m(t_0)|c_{i}^{\dagger}(t-\tau) |\Psi_n(t_0)\rangle\langle \Psi_n(t_0)| c_j(t)|\Psi_m(t_0)\rangle 
 \label{G_less}
\end{equation}
and switching from Heisenberg to Schrödinger picture by transforming the temporal dependence from the operators to the wave-functions we obtain: 
\begin{equation}
G^{<}_{ij}(t,t-\tau)= i \sum_{m,n} \langle \Psi_m(t-\tau)| c_{i}^{\dagger}|\Psi_n(t-\tau) \rangle\langle \Psi_n(t)|c_j|\Psi_m(t)\rangle,
\label{G_Schroed}
\end{equation}
where $|\Psi_m(t)\rangle$ are time-dependent basis states evolving from the field free states under the influence of the full propagator:
$|\Psi_m(t)\rangle =\mathcal{T}e^{-i\int_0^{t}H(t')dt'}|\Psi_m(t_0)\rangle$, and $\mathcal{T}$ is time-ordering operator. Since the evolution is unitary the time-dependent basis states remain orthogonal to each other at any time $t$ within the pulse. 

Since our Hamiltonian explicitly depends on $t$ and $t_{CEP}$, $H(f(t),t+t_{CEP})$, where $f(t)$ is a short pulse envelop, the derivatives of the Hamiltonian with respect to each of this times can be explicitly  calculated:
\begin{equation}
\frac{\partial H}{\partial{t}}=\frac{\partial H}{\partial{f}}\frac{\partial f}{\partial{t}}+\frac{\partial H}{\partial{t_{CEP}}}.
\label{Ham_der}
\end{equation}
yielding
\begin{equation}
 \Delta H=\frac{\partial H}{\partial{f}}\frac{\partial f}{\partial{t}},
\label{Ham_der_dleta}
\end{equation}
where we have introduced an operator $\Delta \equiv \frac{\partial }{\partial{t}}-\frac{\partial }{\partial{t_{CEP}}}$.
To evaluate  $\Delta G^{<}_{ij}(t,t-\tau)$  we take into account that one can rewrite the Schrödinger equation for $\Delta \Psi_m$  in the equivalent form connecting $\Delta H$ and $\Delta \Psi_m$ explicitly ($|\Psi_m(t_0)\rangle$=0):
\begin{equation}
\left|\Delta \Psi_m(t)\right\rangle=-i\int_{t_0}^{t}dt'\mathcal{T}e^{-i\int_{t'}^{t}H(t'')dt''}\Delta H \mathcal{T}e^{-i\int_{t_0}^{t'}H(t'')dt''}|\Psi_m(t_0)\rangle.
\label{WF_der_Delta}
\end{equation}
%\begin{equation}
%\left|\frac{\partial\Psi(t-\tau)}{\partial{t}}\right\rangle=-i\int_{t_0}^{t-\tau}dt'\mathcal{T}e^{-i\int_{t'}^{t-\tau}H(t'')dt''}\frac{\partial H(t')}{\partial{t}}\mathcal{T}e^{-i\int_{t_0}^{t'}H(t'')dt''}|\Psi_m(t_0)\rangle.
%\label{WF_der_t}
%\end{equation}

%Using Eq. (\ref{Ham_der}) we can establish the connection between the respective derivatives of the wave-functions in Eqs.(\ref{WF_der_CEP},\ref{WF_der_t}):
%\begin{equation}
%|\Delta \Psi(t)\rangle=\left|\frac{\partial\Psi(t)}{\partial{t}}\right\rangle-\left|\frac{\partial\Psi(t)}{\partial{t_{CEP}}}\right\rangle=-i\int_{t_0}^{t}dt'\mathcal{T}e^{-i\int_{t'}^{t}H(t'')dt''}\frac{\partial H}{\partial{f}}\frac{\partial f}{\partial{t'}}\mathcal{T}e^{-i\int_{t_0}^{t'}H(t'')dt''}|\Psi_m(t_0)\rangle,
%\label{WF_der_diff}
%\end{equation}
Writing $\Delta G^{<}_{ij}$ explicitly we obtain:
\begin{eqnarray}
\Delta G^{<}_{ij}=i \sum_{m,n}  \langle \Delta\Psi_m(t-\tau)| c_{i}^{\dagger}|\Psi_p(t-\tau) \rangle\langle \Psi_n(t)|c_j|\Psi_m(t)\rangle+\\
i \sum_{m,n} \langle \Psi_m(t-\tau)| c_{i}^{\dagger}|\Delta\Psi_p(t-\tau) \rangle\langle \Psi_n(t)|c_j|\Psi_m(t)\rangle +\\i \sum_{m,n} \langle \Psi_m(t-\tau)| c_{i}^{\dagger}|\Psi_n(t-\tau) \rangle\langle \Delta\Psi_n(t)|c_j|\Psi_m(t)\rangle +\\ i \sum_{m,n}\langle \Psi_m(t-\tau)| c_{i}^{\dagger}|\Psi_n(t-\tau) \rangle\langle \Psi_n(t)|c_j|\Delta\Psi_m(t)\rangle.
\end{eqnarray}
Substituting Eq. \ref{WF_der_Delta} into above equations and limiting ourselves to the terms of the leading order with respect to $\frac{\partial }{\partial{t}}$ and $\frac{\partial }{\partial{f}}$, we find that $\Delta G^{<}_{ij}$ is proportional to the amplitudes of non-adiabatic transitions $a_{mp}(t)=\langle\Psi_m(t) |\frac{\partial }{\partial{f}}|\Psi_p(t)\rangle$ and  between the quasienergy states:
\begin{eqnarray}
\label{Delta_G1}
\Delta G^{<}_{ij}=-\sum_{m,p} f(t)a_{mp}(t)  \langle \Psi_p(t_0)|\bigg[c_j(t),H^{H}(t)\bigg] c_{i}^{\dagger}(t-\tau)|\Psi_m(t_0) \rangle+\nonumber\\+\sum_{m,p} f(t-\tau)a_{mp}(t-\tau) \langle \Psi_p(t_0)| \bigg[c_{i}^{\dagger}(t-\tau),H^{H}(t-\tau)\bigg] c_{j}(t)|\Psi_m(t_0) \rangle,
\label{Delta_G2}
\end{eqnarray}
where
$H^{H}(t-\tau)$ is the Hamilton operator in the Heisenberg picture.
Substituting the explicit expressions for the the commutators  $\bigg[c_{i}^{\dagger}(t-\tau),H^{H}(t-\tau)\bigg]$ and $\bigg[c_j(t),H^{H}(t)\bigg]$ for the Hubbard model and focussing on the correlated part of the Hamiltonian $H_2=U\sum_{i}n_{i\sigma}n_{i\sigma}$:
\begin{eqnarray}
\bigg[c_{i}^{\dagger}(t-\tau),H_2^{H}(t-\tau)\bigg]=-U c_{i,\sigma}^{\dagger}(t-\tau)n_{i,\overline{\sigma}}(t-\tau),\\
\bigg[c_j(t),H_2^{H}(t)\bigg]=U c_{i,\sigma}(t)n_{i,\overline{\sigma}}(t).
\end{eqnarray}
we obtain that $\Delta G^{<}_{ij}$ ($\Delta G^{<}_{ij\sigma}$ ) plotted in Figure \ref{Fig3} directly reflects the non-equilibrium two-body Green's functions $K_{ij\sigma}^{pm}(t,t-\tau)$ and $\tilde{K}_{ij\sigma}^{pm}(t,t-\tau)$:
\begin{eqnarray}
\label{Delta_G1}
\Delta G^{<}_{ij\sigma}(t,t-\tau)=-\sum_{m,p} f(t-\tau)a_{mp}(t-\tau)K_{ij\sigma}^{pm}(t,t-\tau) -\sum_{m,p} f(t)a_{mp}(t)\tilde{K}_{ji\sigma}^{pm}(t,t-\tau)\\\nonumber+\Delta G^{(1)<}_{ij\sigma},
\label{Delta_G2}
\end{eqnarray}
where $f(t)$ is the envelop of the short pulse, $a_{mp}(t)=\langle\Psi_m(t) |\frac{\partial }{\partial{f}}|\Psi_p(t)\rangle$ is an amplitude of non-adiabatic transitions between time-dependent states evolving from the field-free eigenstates $|\Psi_m(t_0) \rangle$, $|\Psi_p(t_0) \rangle$, $\Delta G^{(1)<}_{ij}$ are the non-adiabatic terms of one particle nature and 
\begin{eqnarray}
\label{K_nonequilib1M}
  K_{ij\sigma}^{pm}(t,t-\tau)=U\langle \Psi_p(t_0)|  c^{\dagger}_{i\sigma}(t-\tau)n_{i\overline{\sigma}}(t-\tau) c_{j,\sigma}(t)|\Psi_m(t_0) \rangle,\\
   \tilde{K}_{ij\sigma}^{pm}(t,t-\tau)=U\langle \Psi_p(t_0)|  c_{j\sigma}(t)n_{j\overline{\sigma}}(t) c^{\dagger}_{i,\sigma}(t-\tau)|\Psi_m(t_0) \rangle.
   \label{K_nonequilib2M}
\end{eqnarray}
%\label{K_nonequilib}
 %K_{ij\sigma}^{pm}(t,t-\tau)=U\langle \Psi_p(t_0)|  c^{\dagger}_{i\sigma}(t-\tau)n_{i\overline{\sigma}}(t-\tau) c_{j,\sigma}(t)|\Psi_m(t_0) \rangle,
%\end{equation}
Here $U$ is the on-site Coulomb interaction,
$c_{i\sigma}^{\dagger}$ ($c_{j\sigma}$) are the fermionic creation (annihilation) operators for site $i$ ($j$) and spin $\sigma$. 
$n_{i\sigma}=c_{i\sigma}^{\dagger}c_{i\sigma}$ is the 
particle number operator. Note that in contrast with the standard equilibrium (diagonal in $m$)  expressions:
\begin{eqnarray}
\label{K_equilib1}
 K_{ij\sigma}^{mm}(t,t-\tau)=U\langle \Psi_m(t_0)|  c^{\dagger}_{i\sigma}(t-\tau)n_{i\overline{\sigma}}(t-\tau) c_{j,\sigma}(t)|\Psi_m(t_0) \rangle,\\
 \tilde{K}_{ij\sigma}^{mm}(t,t-\tau)=U\langle \Psi_m(t_0)|c_{j\sigma}(t)n_{j\overline{\sigma}}(t) c^{\dagger}_{i,\sigma}(t-\tau)|\Psi_m(t_0) \rangle,
 \label{K_equilib2}
\end{eqnarray}
%\begin{equation}
%\label{K_equilib}
 %K_{ij\sigma}^{mm}(t,t-\tau)=U\langle \Psi_m(t_0)|  c^{\dagger}_{i\sigma}(t-\tau)n_{i\overline{\sigma}}(t-\tau) c_{j,\sigma}(t)|\Psi_m(t_0) \rangle.
%\end{equation}
 which would appear in ${\partial G_{ij\sigma}}/{\partial t}$, off-diagonal $K_{ij\sigma}^{pm}(t,t-\tau)$ and $\tilde{K}_{ij\sigma}^{pm}(t,t-\tau)$ in Eqs.(\ref{K_nonequilib1M}, \ref{K_nonequilib2M}) feature non-equilibrium two-particle correlations, which would not be recorded in the long pulses ($\Delta G^{<}_{ij}= 0$), corresponding to the standard regime of cycle averaged field driven dynamics, which lays at the foundations of the Floquet engineering. 

\subsection{\label{recovering}Recovering full Green's function from photo-electron measurements.}

While time and angular-resolved photoemission (trARPES) experiments are directly related to the Green's function,  going back from trARPES to the Green's function is  nontrivial. In particular, analyzing the photoemission from a single pulse is restricted by energy-time uncertainty \cite{freericks2009theoretical}. The multi-pulse spectroscopy  does not suffer from this limitation, and allows for full retrieval of the Green function, see e.g \cite{randi2017bypassing} and discussion below.

The full information in the Green's function can be retrieved by suitable measurements, e.g., exploiting the dependence of the photoemission signal on the phase delay between interfering  parts of the photoemission pulse \cite{randi2017bypassing}. With this in mind, we can say that $G^<(t,t-\tau)$  emulates the photoionization signal arising from the interference of two photo-ionization events at $t$ and $t-\tau$.%, which we will refer to as the $P1$ and $P2$ pulse in the following.

To demonstrate how a time-resolved photo-emission experiment may in principle resolve the full Green's function, we start from the general expression given in Ref.~\cite{freericks2009theoretical},
\begin{align}
I(\omega,t_p) 
= 
\int \!\!dt\,dt'\,\,
e^{i\omega(t-t')} 
(-i)G
^{<}
(t,\!t')\,\,s(t)s(t')^*,
\label{pesq}
\end{align}
where orbital and momentum indices are omitted form simplicity, $\omega$ is the frequency of a probe pulse, and $s(t)$ its envelope. It is easy to see that a single Gaussian probe pulse of width $\delta t$ implies a measurement of $G^<(\omega,t)$ with an uncertainty-limited filter in time and frequency. However, with suitable pulses, Eq.~\eqref{pesq} shows that in principle the full time dependence can be retrieved from experiment. For example, to measure $G^<(t,t')$ in a given time window, we choose an orthonormal basis $\phi_n(t)$ for time-dependent functions in that interval, and expand $-iG^<(t,t')=\sum_{n,n'} \phi_n^*(t) g_{n,n'}\phi_{n'}(t')$. The matrix $g_{n,n'}$ is hermitian and positive definite. A probe pulse $S(t)=\phi_{n}(t)$ then measures the diagonal components, $I^< = g_{n,n}$. A probe pulse $S(t)=\phi_n(t)+e^{i\varphi}\phi_{m}(t)$ gives $I^<=g_{n,n}+g_{m,m} + e^{-i\varphi}g_{n,m}+e^{i\varphi}g_{n,m}$, so that off-diagonal components $g_{n,m}$ can be obtained by scanning the phase difference $\varphi$.

\subsection{Spectrocopic nature of the double-time lesser Green's function}
The double-time lesser Green's function provides information about the spectrum of occupied states of the system and is indispensable  for visualizing electronic structure and dynamics.  We review the emergence of laser-dressed states in its structure using the approach presented in section "Direct access to non-adiabatic many body dynamics" starting from 
 Eq. (\ref{G_Schroed}).
Consider the typical Floquet regime corresponding to CW pulse. The quasienergy  states can be written as:
\begin{equation}
    \Psi_m(t)=e^{-i\mathcal{E}_mt}f_m(t),
    \Psi_n(t)=e^{-i\mathcal{E}_nt}f_n(t),
\end{equation}
where $\mathcal{E}_m$ ($\mathcal{E}_n$) is the quasienergy of a Floquet state $m$ ($n$) and $f_m(t)$ ($f_n(t)$) is a periodic function of time.
Introducing auxiliary functions $\Phi_{nm}(t)$, $\Phi^{(+)}_{mn}(t-\tau)$ 
\begin{equation}
    \Phi_{nm}(t)\equiv\langle \Psi_n(t)|c_j|\Psi_m(t)\rangle= e^{i(\mathcal{E}_n-\mathcal{E}_m)t}f_{nm}(t),
\end{equation}
\begin{equation}
    \Phi^{(+)}_{mn}(t-\tau)\equiv\langle \Psi_m(t-\tau)| c_{i}^{\dagger}|\Psi_n(t-\tau) \rangle= e^{-i\mathcal{E}_n(t-\tau)+i\mathcal{E}_m(t-\tau)}f_{mn}(t-\tau),
\end{equation}
we can rewrite the expression for Green's function
\begin{equation}
G^{<}_{ij}(t,t-\tau)= i \sum_{m,n}\Phi^{(+)}_{mn}(t-\tau)\Phi_{nm}(t)=\sum_{m,n}f_{mn}(t-\tau)f_{nm}(t)e^{-i(\mathcal{E}_m-\mathcal{E}_n)\tau}
\label{G_lesser_Phi}
\end{equation}
Since the functions $f_{mn}(t),f_{nm}(t-\tau)$ are periodic, we can expand them in Fourier series:
\begin{eqnarray}
f_{nm}(t)=\sum_{k}a^{nm}_{k} e^{i\omega kt},\\
f_{mn}(t-\tau)=\sum_{k'}a^{mn}_{k'} e^{-i\omega k'(t-\tau)}.
\end{eqnarray}
Thus, 
\begin{equation}
G^{<}_{ij}(t,t-\tau)= i \sum_{m,n} \sum_{k',l}a^{mn}_{k'}a^{nm}_{k'+l} e^{-il\omega t} e^{-i(\mathcal{E}_m-\mathcal{E}_n+k'\omega)\tau}
\label{G_less_before_FT}
\end{equation}
and the Fourier transform wrt $\tau$ yields $\Omega_{\tau}=\mathcal{E}_m-\mathcal{E}_n+k'\omega$, where $\mathcal{E}_m-\mathcal{E}_n$ represent spectral energies on the vertical axis of Fig \ref{Fig1}(a,c,e). If we fix $\Omega_{\tau}=-1.25$ eV ($LHB$), then $k'=0$. Fourier transforming Eq. \ref{G_less_before_FT} wrt $t$ we obtain $\Omega_t=l\omega$ and $a^{nm}_{l}$ are the amplitudes of the Floquet ladder starting from zero energy, i.e. the Floquet ladder corresponding to $LHB$ for $\Omega_{\tau}=-1.25$ eV.  Thus, in the standard Floquet picture, fixing  $\Omega_{\tau}=E$ (as it is done in Fig\ref{Fig3}) leads to observation of  standard Floquet ladder from the state $E$. 

The situation changes dramatically in the presence of non-adiabatic transitions between the Floquet states. Suppose such non-adiabatic transition couples the  Floquet state $m$ to another  Floquet state $m'$. Then
\begin{equation}
    \Phi_{nm}(t)=\lambda_{mm}(t) e^{i(\mathcal{E}_n-\mathcal{E}_m)t}f_{nm}(t)+\lambda_{m'm}(t) e^{i(\mathcal{E}_n-\mathcal{E}_{m'})t}f_{nm'}(t),
\end{equation}
and
\begin{equation}
    \Phi^{(+)}_{mn}(t-\tau)=\lambda_{mm}(t-\tau) e^{-i(\mathcal{E}_n-\mathcal{E}_m)(t-\tau)}f_{mn}(t-\tau)+\lambda_{mm'}(t-\tau) e^{-i(\mathcal{E}_n-\mathcal{E}_{m'})(t-\tau)}f_{m'n}(t-\tau),
\end{equation}
%\equiv \sum_{m,n} G^{<nm}_{ij}(t,t-\tau)
where the coefficients $\lambda_{mm'}(t)$ represent the amplitudes of non-adiabatic transitions between the quasienergies $\mathcal{E}_m$ and $\mathcal{E}_{m'}$.
In this case, the product $\Phi_{mn}(t-\tau)\Phi_{nm}(t)$, contributing to  $G^{<}_{ij}(t,t-\tau)$  in Eq.(\ref{G_lesser_Phi}), acquires three additional terms (Eqs. (\ref{term_2}-\ref{term_4})):
\begin{eqnarray}
\Phi_{mn}(t-\tau)\Phi_{nm}(t)= 
\lambda_{mm}(t)\lambda_{mm}(t-\tau)f_{nm}(t)f_{mn}(t-\tau) e^{i(\mathcal{E}_n-\mathcal{E}_m)\tau}+
\label{term_1}\\
\label{term_2}
\lambda_{m'm}(t)\lambda_{mm}(t-\tau)f_{nm'}(t)f_{mn}(t-\tau) e^{i(\mathcal{E}_{m}-\mathcal{E}_{m'})t} e^{i(\mathcal{E}_n-\mathcal{E}_m)\tau}+
\\
\label{term_3}
\lambda_{mm}(t)\lambda_{mm'}(t-\tau)f_{nm}(t)f_{m'n}(t-\tau) e^{i(\mathcal{E}_{m'}-\mathcal{E}_m)t} e^{i(\mathcal{E}_n-\mathcal{E}_{m'})\tau}+\\\label{term_4}
\lambda_{m'm}(t)\lambda_{mm'}(t-\tau)f_{nm'}(t)f_{m'n}(t-\tau)  e^{i(\mathcal{E}_n-\mathcal{E}_{m'})\tau}.
\label{G_less_before_FT_NA}
\end{eqnarray}
Non-adiabatic transitions from state with quasienergy $\mathcal{E}_{m}$ to state with  quasienergy $\mathcal{E}_{m'}$  lead to new features in the spectrum both along $\Omega_{\tau}$ and $\Omega_{t}$ dimensions. Indeed, while the term Eq.(\ref{term_1}) is similar to  Eq. (\ref{G_lesser_Phi}), the term Eq.(\ref{term_4}) adds new  frequency to the spectrum along $\Omega_{\tau}$ dimension.  In addition, 
terms represented by Eqs. (\ref{term_2},\ref{term_3})  oscillate (in time $t$) at "new" frequencies $\pm(\mathcal{E}_{m'}-\mathcal{E}_m)$, which will appear along $\Omega_{t}$ direction. Thus, we see that non-adiabatic transitions lead to significant restructuring of the spectrum encoded in $G^{<}_{ij}(t,t-\tau)$ and revealed after Fourier transforms wrt to $\tau$ and $t$.

To reveal the time-scale of non-adiabatic transitions we need to employ the additional dimension, sensitive to sub-laser cycle features  of electron dynamics.
The carrier-envelope phase (CEP) is a natural choice.
Scanning CEP we obtain $G^{<}_{ij}(t,t-\tau,t_{CEP})$.
Fourier transform wrt $t_{CEP}$\cite{PhysRevLett.99.220406} allows us to tag each non-adiabatic transition and quantify the role of sub-laser cycle dynamics in restructuring the spectrum of the system and in the formation of the final insulating state.

\section{\label{Supplementary Information}Supplementary Information}

\subsection{\label{Floquet} Cycle-averaged Floquet  picture}

\begin{figure}[h!]
\begin{minipage}[h]{0.49\linewidth}
\begin{overpic}[width=1\textwidth]{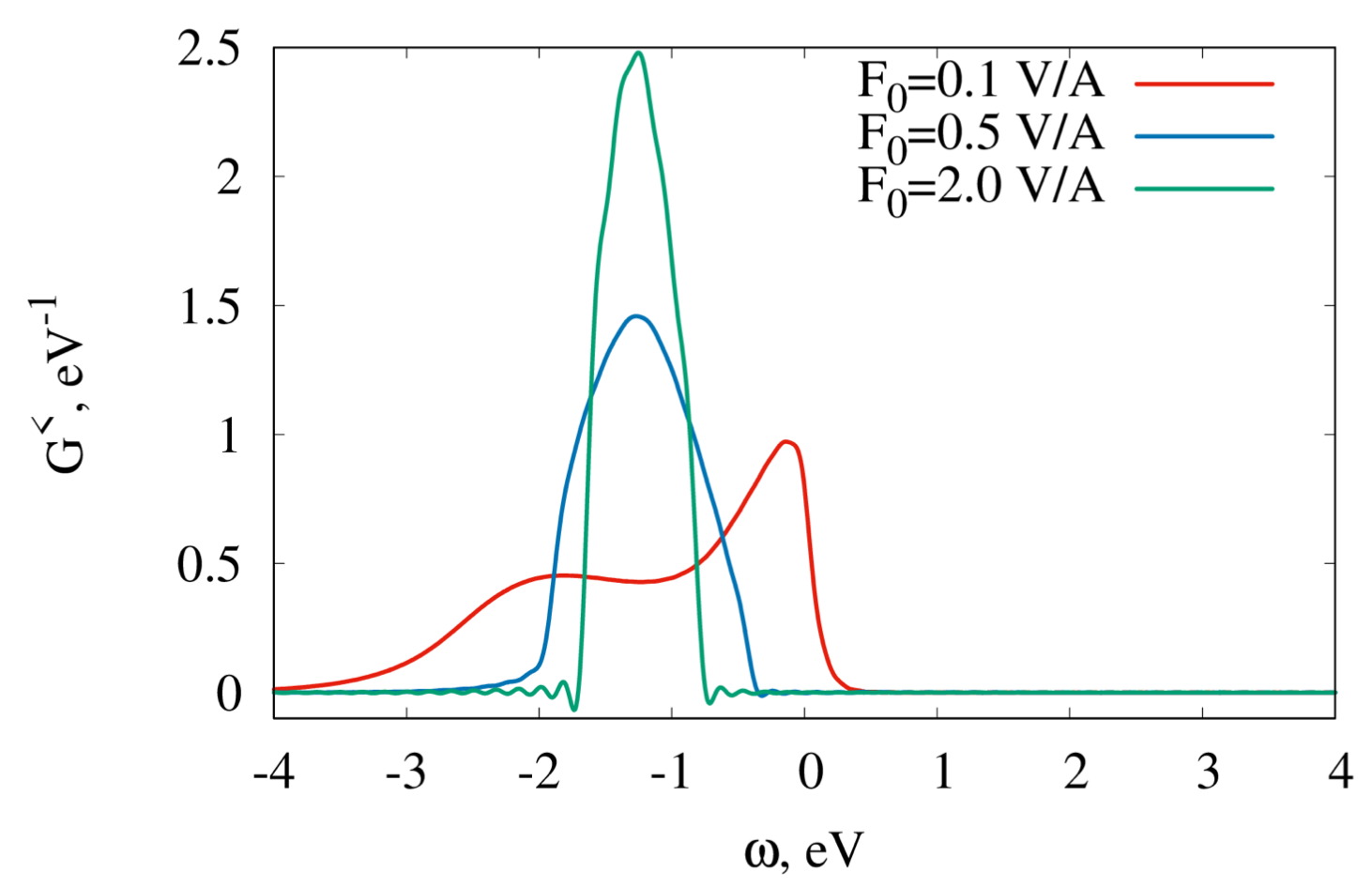}
 \put (22.5,56) {\textcolor{black}{(a)}}
\end{overpic}
\end{minipage}
\hfill
\begin{minipage}[h]{0.49\linewidth}
\begin{overpic}[width=1.2\textwidth]{figures_/Fig_1/1a.png}
 \put (22.5,50) {\textcolor{white}{(b)}}
\end{overpic}
\end{minipage}

\caption{(a) Equilibrium density of occupied states calculated with renormalized hoppings $T_{ij}^{eff}=T_{ij} J_0(\textbf{A}\textbf{R}_{ij})$ and the same $U$ (i.e. $U=2.5$ eV). Renormalization yields: $T_{ij}^{eff}=T_{ij} J_0 (0.32)=0.9745T_{ij}$ (for $F_0=0.1$ V/A); 
$T_{ij}^{eff}=T_{ij} J_0 (1.6)=0.45T_{ij}$ (for $F_0=0.5$ V/A); 
$T_{ij}^{eff}=T_{ij} J_0 (6.4)=0.24T_{ij}$ (for $F_0=2$ V/A.) (b) Fig.\ref{Fig1}a of the paper reflecting temporal evolution of density of states for $F_0=0.1$ V/A.
}
\label{Fig5}
\end{figure}

Here we show that the metal-insulator transition we find in our simulations can not be understood in terms of the simple cycle-averaged picture, which underlies the standard mechanism of  Floquet engineering. It becomes clear  after the analysis of results of laser-driven dynamics at the  lowest field strength $F_0=0.1$ V/A, shown in Fig.\ref{Fig5}.
%The lowest field $F_0=0.1$ V/A already emphasizes the  importance of sub-cycle dynamics. 
Comparing the prediction based on the cycle averaged picture show in in Fig.\ref{Fig5}(a), which uses renormalized hoppings, with the full simulation shown in Fig.\ref{Fig5}(b), we see that panel (a) does not describe our observations. 
Specifically, the red curve in Fig.\ref{Fig5}(a) presents equilibrium density of occupied states calculated with renormalized hoppings $T_{ij}^{eff}=T_{ij} J_0(\textbf{A}\textbf{R}_{ij})$ and the same $U$ (i.e. $U=2.5$ eV). It is  essentially identical to the electron density prior to the pulse,  because for $F_0=0.1$ V/A the Bessel function  $J_0\simeq 1$, meaning that the hoppings are hardly  modified.
In Fig.\ref{Fig5}(b) (identical to  Fig.\ref{Fig1}a) we see  that at about 17.5 fs the peak of the density is shifted towards  the energy around -1.25 eV.  Remarkably it stays there after the field is off. The cycle averaged picture  ( panel (a), red curve) suggests that the peak of the density should be around zero energy at all times.  
The stark contrast between the density in  Fig.\ref{Fig5}a and the density on in Fig.\ref{Fig5}b  after 17.5 fs means that the sub-cycle modification of the electron density is crucial for establishing the state of the system observed after 17.5 fs and after the pulse is off.

\subsection{\label{Locking}Temporal dynamics in the locking regime}
We have established three different regimes of electron dynamics, described in Fig. \ref{Fig1} (b,d,f). Here we focus on the most interesting regime prominent at  $F_0=2$ V/A, which corresponds to pronounced locked oscillations of electron densities at  LHB (red) and UHB (blue) energies: after 13 fs the populations at these energies  are nearly equal, locked in phase, and oscillate out of phase with  the electron density at the QP energy (green). Here we provide additional information to show that these oscillations are well synchronized with the instantaneous electric field.

\begin{figure}[h!]
\begin{minipage}[h]{0.49\linewidth}
\begin{overpic}[width=1\textwidth]{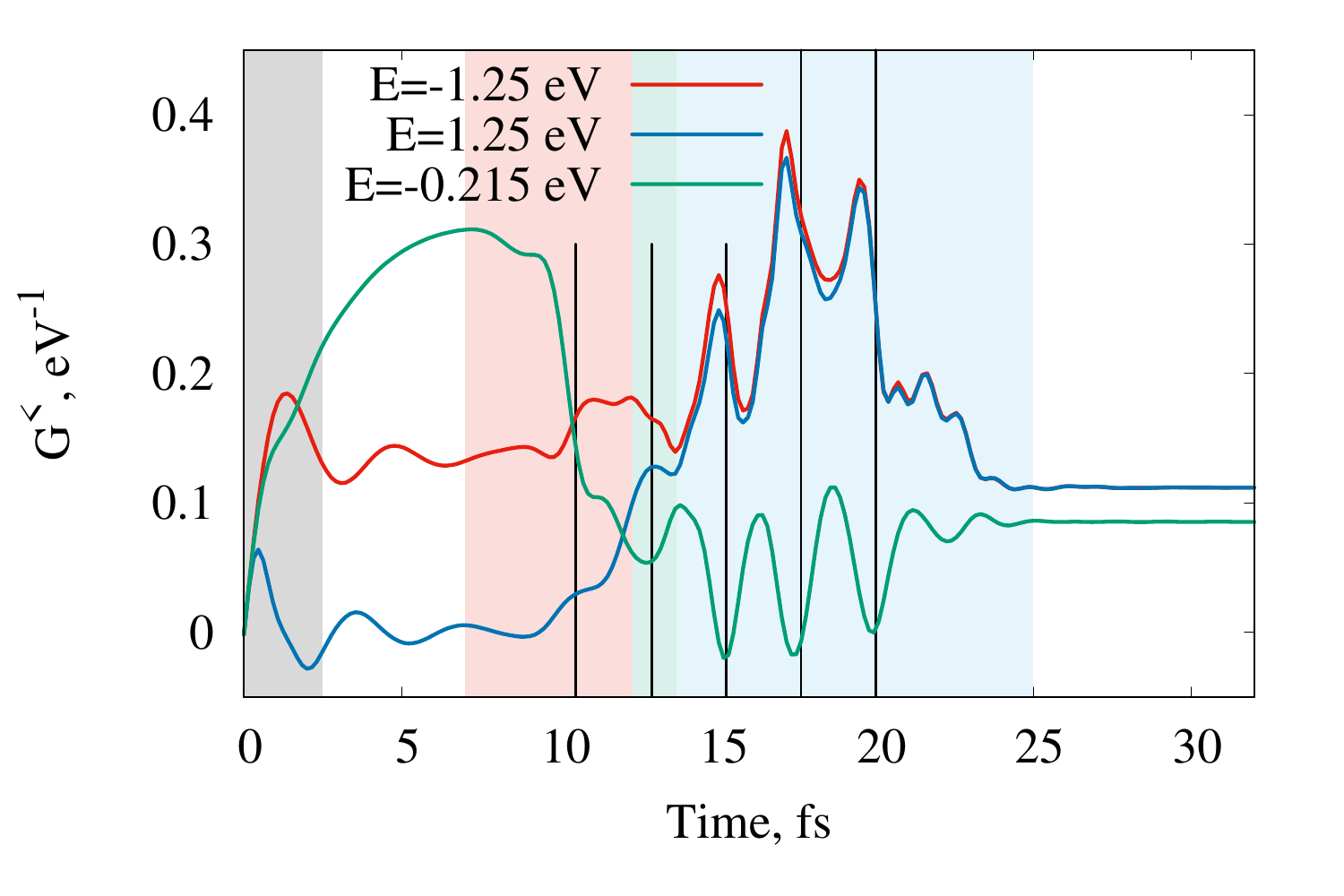}
 \put (20.5,58) {\textcolor{black}{(a)}}
\end{overpic}
\end{minipage}
\hfill
\begin{minipage}[h]{0.49\linewidth}
\begin{overpic}[width=1\textwidth]{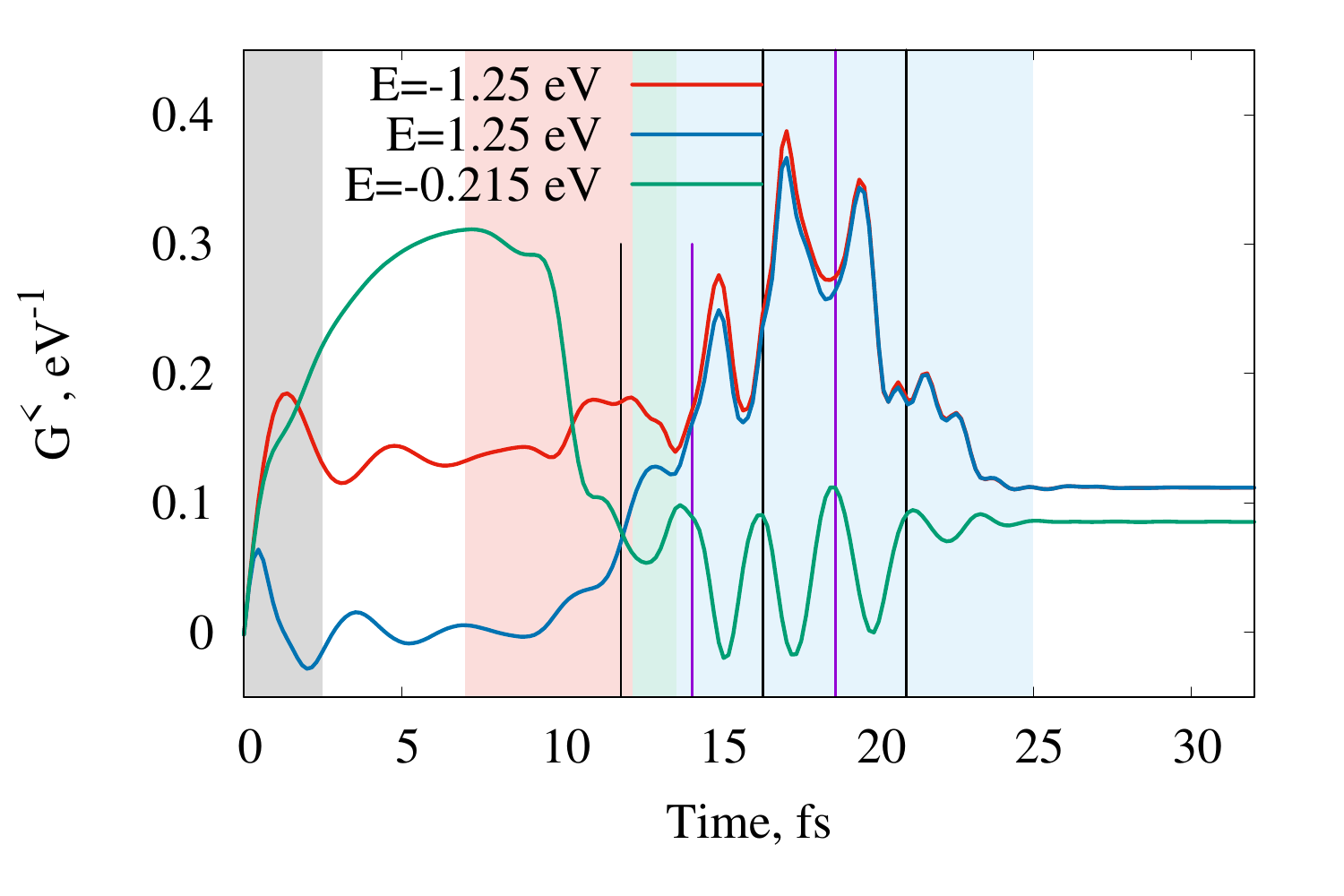}
 \put (20.5,58) {\textcolor{black}{(b)}}
\end{overpic}
\end{minipage}

\caption{Temporal oscillations of electron density at the key energies of the system for $F_0=2$ V/A, shown vs oscillations of the laser electric field.  (a) The vertical lines mark  time instants, at which the instantaneous field is equal to zero, $F(t)=0$. (b) Vertical lines mark  time instants, at which the instantaneous field is maximal, $|F(t)|=F_0$.
}
\label{Fig6}
\end{figure}

Fig .\ref{Fig6} (a) specifies the time instants at which the instantaneous field is $F(t)=0$. It shows that the populations at LHB (red) and  UHB (blue) reach maxima just before the instantaneous zero of the field. The instantaneous rate of population decay (time derivative of red (LHB)  and blue (UHB) curves) appears to be maximized at $F(t)=0$. Thus, the rate of flow of electron density \textit{from} $LHB$ and $UHB$ bands maximizes near zeroes of field oscillation ($F(t)\simeq 0$).
The population at  the QP (green) reaches minima at the zeroes of the field, meaning that it is in phase with the oscillations of the laser field.

Fig.\ref{Fig6} (b) specifies time instants at which the instantaneous field reaches its maximal value $|F(t)|=F_0$. It shows that  the  populations  in LHB (red) and UHB (blue) reach minima just before the maximum of the field, when the electron density at the energy of $QP$ maximises. The instantaneous rate of the population increase (time derivative of red (LHB)  and blue (UHB) curves), appears to be maximized near the  maxima of the laser field oscillation ($|F(t)|\simeq F_0$).

Thus, the charge density oscillates between $QP$ and  $LHB$,  $UHB$ and  these oscillations are synchronized with the instantaneous field. Figs. \ref{Fig5} and \ref{Fig6} emphasize the importance of the sub-laser cycle dynamics and motivate the development of the sub-cycle multidimensional spectroscopy described in the main text. This spectroscopy allows one to zoom into the sub-cycle electron dynamics and identify dominant pathways of charge flow responsible for the metal-insulator transition in our system.   

\subsection{\label{Benchmark}Benchmark simulations}
\begin{figure}[h!]
 \includegraphics[width=1\linewidth,angle=0]{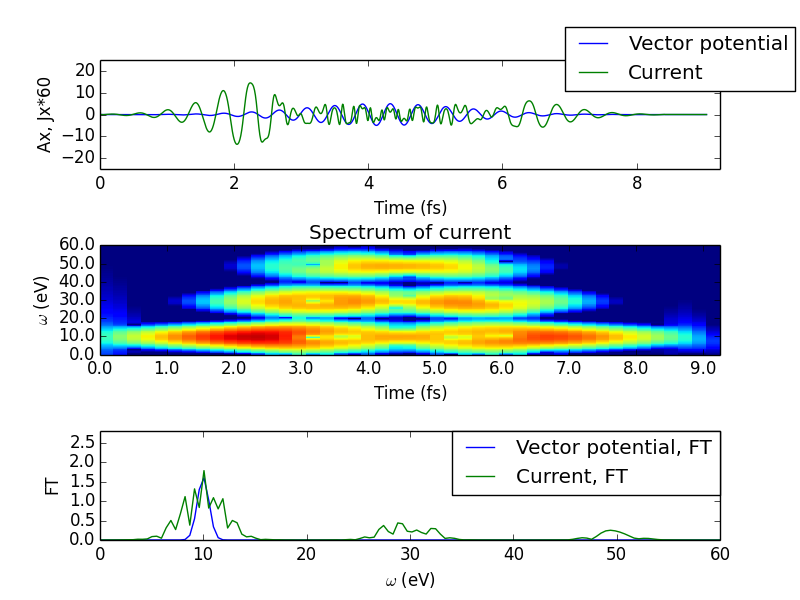}
\caption{High harmonic generation on square lattice for gaussian pulse with central frequency $\omega=10$ eV, $FWHM=3$ fs, and $A_0=5$, polarization along [10] direction. Upper panel: vector potential (blue) and current (green); middle panel: Gabor transform of the current; Lower panel: spectra of incoming pulse (blue) and current (green).}
\label{HHGtestVitja}  
\end{figure}
\begin{figure}[h!]
 \includegraphics[width=1\linewidth,angle=0]{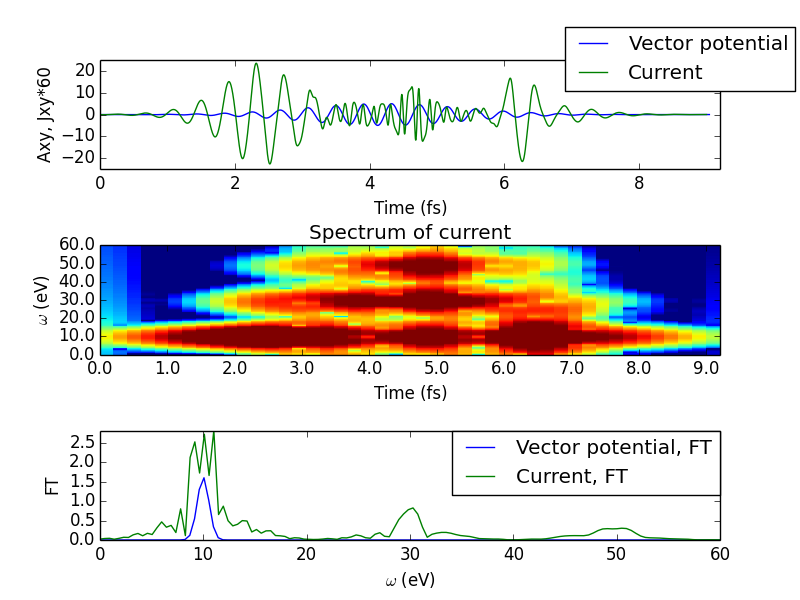}
\caption{High harmonic generation on 12-sites chain for Gaussian pulse with central frequency $\omega=10$~eV, $FWHM=3$~fs, and $A_{0}=5$, polarization along the chain. Upper panel: vector potential (blue) and current (green); middle panel: Gabor transform of the current (in the ground state); Lower panel: spectra of incoming pulse (blue) and current (green).
%, red, cyan). 
%[@HA: I see no red and cyan curves]
}
\label{HHGtestRui}  
\end{figure}
We have  benchmarked  our IPT-DMFT on square lattice against the code described in Ref.~\cite{silva2018high}, performing exact diagonalization for the finite 12-site one-dimensional chain.
We set the hoppings $T=1$~eV, on-site Coulomb repulsion $U=6$~eV, pulse vector potential amplitude $A_0=5$, pulse FWHM is 3fs, pulse central frequency $\omega=10$~eV. In order to compare our two-dimensional lattice model to one-dimensional chain, we choose linear pulse polarization along [10] direction and relatively large field amplitude.

Although the physics is different between 1D and 2D systems due to the existence of closed loops and additional scattering channels in two dimensions, the resulting HHG spectra look qualitatively similar (see Figs.~\ref{HHGtestVitja} and \ref{HHGtestRui}). 

\end{document}